\documentclass[preprint,showkeys]{revtex4}
\usepackage{graphicx}
\usepackage{amsmath,amssymb}
\usepackage{tensor}
\usepackage{xcolor}   
\usepackage[font={small}]{caption}

\begin{document}
	
	\title{ Cosmological Tensions with Non-Extensive Entropic Cosmology: A Modified Stress-Energy Approach}
	\author{ A.Khodam-Mohammadi$^{1}$\footnote{Email:khodam@basu.ac.ir} and M. Monshizadeh$^{2}$}
	\affiliation{$^{1}$Department
		of Physics, Faculty of Science, Bu-Ali Sina University, Hamedan
		65178, Iran \\ $^{2}$Department of Physics, Ha.C., Islamic Azad University, Hamedan, Iran} 
	
	\begin{abstract}
	
	We perform a comprehensive cosmographic analysis of Friedmann cosmologies modified by non-extensive entropy frameworks, focusing on Tsallis, R\'enyi, and Kaniadakis entropies within a novel modified energy-momentum tensor approach. In our approach the microscopic matter density remains $\rho=m~n$
    while the horizon thermodynamics of non-extensive entropy modifies the effective source that drives expansion, $\rho_{eff}=f(\rho)\rho$
    which reduces to the standard case for $f(\rho)=1$. By deriving the generalized Friedmann equations for each entropy type, we calculate analytical expressions for key cosmographic parameters, including the deceleration ($q_0$), jerk ($j_0$), snap ($s_0$), and lerk ($l_0$) parameters, and examine their behavior compared to the standard $\Lambda$CDM model.	
	Our results reveal significant differences between the traditional formal approach and the modified energy-momentum approach, particularly in the Tsallis model where cosmographic parameters and the dimensionless Hubble function $E(z)$ show notable deviations for deformation parameters away from unity. Moreover, both R\'enyi and Kaniadakis models exhibit increasing tension with $\Lambda$CDM at higher redshifts, while remaining similar at low redshifts.	
	Importantly, given the persistent $\sim5\sigma$ Hubble tension between local and global measurements, our analysis indicates that the flexibility of the modified entropic cosmology framework to alter the expansion rate could potentially alleviate this discrepancy by modifying the effective expansion history in a way compatible with observations. Future work will involve full Bayesian analyses with Pantheon+, DESI, and Planck data to further assess the viability of these models in resolving current cosmological tensions.

	\end{abstract}
	

\maketitle

\section{Introduction}
Entropy is a fundamental concept in thermodynamics and statistical mechanics, playing a crucial role in understanding the large-scale behavior of the universe. In standard cosmology, the Bekenstein-Hawking entropy (BH entropy) of black hole horizons has been essential in linking gravity \cite{Hawking:1975vcx,Bekenstein:1973ur}, thermodynamics \cite{Jacobson:1995ab}, and quantum theory. However, various observational finding such as the accelerated expansion of the universe, the nature of dark energy \cite{Shahhoseini:2025sgl, Brevik:2024nzf, Sheykhi_2023, P:2022amn, Nojiri_2022, Manoharan:2022qll, Di_Gennaro_2022, Bhattacharjee_2021, Moradpour:2020dfm, DAgostino:2019wko, Saridakis_2018}, and the early inflationary phase \cite{Odintsov_2023, Odintsov:2023rqf, Lambiase:2023ryq, Teimoori:2023hpv, Luciano:2023roh, Khodam-Mohammadi:2024iuo}, have motivated exploration of generalized entropy frameworks beyond the traditional Boltzmann-Gibbs formalism \cite{Akbar:2006kj,Cai:2006rs,Sheykhi:2007zp,Sheykhi:2007gi,Jamil:2009eb,Cai:2009ph,Wang:2009zv,Jamil:2010di,Gim:2014nba,Fan:2014ala,Sanchez:2022xfh}. On the other hand, it is important to note that the universe is a self-gravitating system, and gravitation is a long-range interaction that usually breaks extensivity. For this reason, the standard Boltzmann-Gibbs entropy may not be the correct description and non-extensive entropies should be required in cosmology. This may lead to changes in the matter-energy content of the universe. 

Among these frameworks, Tsallis, R\'enyi, and Kaniadakis entropies have attracted considerable interest. These non-extensive entropy forms provide consistent descriptions of systems with long-range interactions, gravitational clustering, and non-equilibrium conditions, all of which are naturally present in cosmology. When applied to the apparent horizon of the FLRW universe, they lead to modified Friedmann equations that can potentially explain cosmic acceleration without requiring exotic matter fields. Such entropy based extensions have opened promising directions in the study of dark energy, early universe thermodynamics, and the deep connections between gravity, information theory, and holography \cite{Sheykhi:2018dpn, Saridakis:2020cqq, Nojiri:2019skr}.

In particular, Tsallis entropy, introduced by Constantino Tsallis in 1988 \cite{Tsallis:1987eu}, generalizes Boltzmann-Gibbs entropy for systems with non-additive and long-range interactions:
\begin{equation}
S_q = k \frac{1 - \sum_i p_i^q}{q - 1},
\end{equation}
where $q$ is the non-additivity parameter. When applied to gravitational systems, this leads to modified thermodynamic quantities. Later, Tsallis and Cirto \cite{Lyra:1997ggy} proposed a generalized area-entropy relation for black hole horizons:
\begin{equation}
S = g A^\delta, \label{tsal}
\end{equation}
where $\delta$ is the non-extensivity index and $g$ is a constant, recovering the standard Bekenstein-Hawking (BH entropy) result when $\delta = 1$.

R\'enyi entropy, introduced by Alfr\'ed R\'enyi  \cite{Renyi:1960}, is defined as:
\begin{equation}
S_R = \frac{1}{1 - q} \ln \sum_i p_i^q,
\end{equation}
and serves as a bridge between Tsallis entropy and standard thermodynamics. Unlike Tsallis entropy, R\'enyi entropy is additive for independent systems and has applications in black hole thermodynamics, statistical complexity, and quantum information theory.

Kaniadakis entropy \cite{Kaniadakis:2005zk,Drepanou:2021jiv}, which emerges from $\kappa$-deformed statistics in relativistic kinetic theory, is given by:
\begin{equation}
S_\kappa = -\sum_i \frac{p_i^{1+\kappa} - p_i^{1-\kappa}}{2\kappa},
\end{equation}
where $\kappa$ is the deformation parameter. In the limit $\kappa \to 0$, it reduces to the standard Boltzmann-Gibbs form. Its statistical and thermodynamical consistency in high-energy and gravitational systems has been well documented.

These entropy models have been applied across various cosmological epochs. Tsallis entropy has been used in both inflationary models and late-time cosmic acceleration \cite{Jawad:2018frx, Tavayef:2018xwx}. R\'enyi entropy has been applied to unified dark sector cosmologies \cite{Moradpour:2016ubd}, while Kaniadakis entropy has been employed to explain both early- and late-time cosmic dynamics \cite{Luciano:2019cna, Ghaffari:2022tiq}.

Traditionally, the modification of cosmology based on generalized entropies proceeds via application of the first law of thermodynamics to the apparent horizon. In this standard thermodynamic approach-pioneered by Jacobson \cite{Jacobson:1995ab} and later extended by Cai and Akbar \cite{Cai:2005ra} the Friedmann equations emerge from the Clausius relation $\delta Q = T dS$. By adopting generalized entropy forms instead of Bekenstein-Hawking entropy, one obtains modified Friedmann equations, while the continuity equation for cosmic fluids remains unchanged. For example a modified Friedmann equation using Tsallis entropy in a flat FRW universe has been derived \cite{Sheykhi_2018, Odintsov:2022qnn, Nojiri_2022}. Saridakis et al. extended such analyses to various entropy forms and studied their observational constraints \cite{Saridakis:2020cqq}, while Odintsov and Oikonomou explored these modifications within modified gravity and loop quantum cosmology frameworks \cite{Odintsov:2020zct, Nojiri:2019skr}, revealing rich phenomenology at both high and low redshifts.

However, in 2023, we introduced a novel approach that offers a deeper thermodynamic reinterpretation of cosmology \citep{Khodam_Mohammadi_2023}. Instead of modifying only the gravitational (left-hand) side of the Friedmann equations, this approach proposes that generalized entropy corrections modify the energy-momentum tensor itself. Consequently, both the Friedmann equation and the continuity equation are altered self-consistently. This method allows derivation of modified dynamics for any entropy form, including Tsallis, R\'enyi, and Kaniadakis, without assuming the standard fluid conservation law. Building on this, in 2024 we extended the framework to early-universe inflationary dynamics in the context of non-extensive entropies \citep{Khodam-Mohammadi:2024iuo}, demonstrating that such entropy contributions can yield viable slow-roll inflation without requiring special scalar field potentials. Importantly, these models produce observational predictions for the scalar spectral index $n_s$ and tensor-to-scalar ratio $r$ consistent with Planck 2018 bounds, thus providing a unified framework for both inflation and dark energy based on thermodynamics.

Meanwhile, recent cosmological observations have revealed a persistent tension between local measurements of the Hubble constant ($H_0$) from Cepheid-calibrated supernovae (e.g. SH0ES) and global determinations from the CMB within $\Lambda$CDM, now exceeding the $\sim 5\sigma$ level \cite{scolnic2025hubble, riess2024jwst, hart2020updated}. This discrepancy has motivated exploration of modified gravity, early dark energy, and alternative thermodynamic or entropic frameworks as potential solutions \cite{vacher2022constraints, tyagi2025constraints}. Non-extensive entropic cosmologies with modified Friedmann dynamics may provide a viable resolution mechanism by altering the expansion history in a way compatible with both early- and late-universe constraints.

These developments motivate further investigation into the cosmological implications of entropy-modified gravity, including their effects on cosmic kinematics.

In this work, we apply a cosmographic approach. Cosmography describes the kinematics of the universe's expansion in a model-independent way using parameters such as the Hubble constant $H_0$, deceleration parameter $q_0$, jerk $j_0$, snap $s_0$, and lerk $l_0$. The Pad\'e approximation allows greater flexibility and accuracy, especially when fitting observational data such as Type Ia supernovae, baryon acoustic oscillations (BAO), and cosmic chronometers, without assuming any specific dark energy or gravity model. Recent studies have demonstrated the effectiveness of Pad\'e-based cosmography. It showed that Pad\'e approximants yield better fits to cosmological distances than Taylor expansions, especially at $z > 1$ \cite{Cattoen:2007sk,Capozziello:2011tj,Rezaei:2020lfy,Li:2019qic,Capozziello:2020ctn,Vitagliano:2009et} and also Pad\'e cosmography reduces error propagation and improves parameter estimation stability \cite{Shahhoseini:2025pyr, rezaei2017constraints}.

In this work, we examine whether our non-extensive entropic models can reproduce the observed expansion history of the universe and whether they can be distinguished from the $\Lambda$CDM model using cosmography. 

This paper is organized as follows. After the introduction, we provide a brief review of the entropic gravity formalism based on two independent approaches for deriving modified Friedmann equations in Sec. \ref{sec2}. In Sec. \ref{sec3}, we present the fundamentals of the cosmography approach, and in Sec. \ref{sec4}, we derive the cosmographic parameters within the new entropic framework. Sec. \ref{sec5} contains the analysis of our results, and finally, we conclude with a summary and discussion of future prospects. Throughout this work we are using Plank mass unit, where `$c=\hbar=G=1$'.

\section{Non-Extensive Entropy Models: A Contrast of Two Approaches}\label{sec2}

In the homogeneous and isotropic FLRW universe, the metric is
\begin{equation}
ds^2 = h_{ab} , dx^a dx^b + R^2 (d\theta^2 + \sin^2\theta , d\phi^2),
\end{equation}
where the coordinates are $x^0 = t$, $x^1 = r$, and $R = a(t)r$. The two-dimensional metric is $h_{ab} = \text{diag}(-1, a^2/(1 - kr^2))$, with determinant $h$.

The apparent horizon plays a key role in cosmological thermodynamics. Relevant quantities include \cite{Sanchez:2022xfh,Cai:2005ra}:
\begin{eqnarray}
S_{BH} &=& \frac{A}{4} = \pi R_h^2 = \frac{3}{8\rho}, \\
R_h^2 &=& \frac{1}{H^2 + \frac{k}{a^2}}, \\
\kappa &=& \frac{1}{2\sqrt{-h}} \frac{\partial}{\partial x^a} \left( \sqrt{-h} , h^{ab} \frac{\partial R_h}{\partial x^b} \right), \\
T_h &=& \frac{|\kappa|}{2\pi} = \frac{1}{2\pi R_h} \left| 1 - \frac{\dot{R}_h}{2 H R_h} \right|.
\end{eqnarray}

Here, the difference between the two approaches emerges.

\subsection{The old (formal) approach}

In the standard method, the energy-momentum tensor of a cosmic perfect fluid is given by
\begin{equation}
T^\mu_\nu = \text{diag}(-\rho, p, p, p),
\end{equation}
where $\rho$ is the energy density and $p$ is the pressure of the cosmic fluid. Conservation of energy-momentum yields
\begin{equation}
\dot{\rho_i} + 3H\rho_i(1 + w_i) = 0,
\end{equation}
where the subscript `$i$' denotes any matter content (dark matter, dark energy, radiation, etc). Applying the first law of thermodynamics on the apparent horizon $R_h$,
\begin{eqnarray}
dE &=& -T dS + W dV, \notag \\
E &=& \frac{4\pi}{3} R_h^3 \rho, \quad V = \frac{4\pi}{3} R_h^3, \quad W = -\frac{1}{2} (\rho - p),
\end{eqnarray}
leads to the modified Friedmann equation. For Tsallis entropy, $S = \gamma A^{\delta}$, where $A$ is the apparent horizon area and $\delta$ is the Tsallis parameter \citep{Sheykhi:2018dpn, Odintsov:2020zct}:
\begin{equation}
\left( H^2 + \frac{k}{a^2} \right)^{2-\delta} = \frac{8\pi}{3} \rho,
\end{equation}
or in comparison to standard Friedmann equation, it gives
\begin{equation}
 H^2 + \frac{k}{a^2} = (\frac{8\pi}{3} \rho)\left(\frac{8\pi}{3} \rho\right)^{\frac{\delta-1}{2-\delta}}.\label{fr-thermo}
\end{equation}

This thermodynamic approach has been fully developed for generalized entropies, but exact solutions to the Friedmann equations remain limited to the Tsallis and Barrow entropies.
\subsection{New approach: modification of the energy-momentum tensor}

In our new approach, we propose a modified energy-momentum tensor to describe deviations from the standard cosmic fluid \citep{Khodam_Mohammadi_2023}:
\begin{equation}
T^\mu_\nu = \text{diag}(-\rho f(\rho), p f(\rho), p f(\rho), p f(\rho)),
\end{equation}
where $f(\rho)$ modifies the perfect fluid without changing the equation-of-state parameter $w = p/\rho$. The conservation law $T^{\mu\nu}_{;\nu} = 0$ then yields:
\begin{equation}\label{consf}
\dot{\rho_i} \left[ 1 + \rho_i \frac{f'(\rho_i)}{f(\rho_i)} \right] + 3H\rho_i(1 + w_i) = 0,
\end{equation}
where $w_m = 0$ (dark matter), $w_d = -1$ (dark energy), and $w_r = 1/3$ (radiation).

Substituting this modified tensor into the Einstein field equations within the FLRW cosmology gives the modified Friedmann equations:
\begin{eqnarray}\label{FR}
H^2 + \frac{k}{a^2} &=& \frac{8\pi}{3} \rho f(\rho), \\
\dot{H} - \frac{k}{a^2} &=& -4\pi (\rho + p) f(\rho).
\end{eqnarray}

Applying the first law of thermodynamics, $dE = -\delta Q + W dV$, with $\delta Q = T dS$, we obtain:
\begin{equation}
E = \frac{4\pi}{3} R_h^3 \rho f(\rho), \quad V = \frac{4\pi}{3} R_h^3, \quad W = -\frac{1}{2} (\rho - p) f(\rho).
\end{equation}
After solving differential equation of the first law, by considering any non-extensive entropy, we obtain \cite{Khodam_Mohammadi_2023}
\begin{equation}
\tilde{S} = S_g(S),
\end{equation}
where $S = \pi R_h^2 = \frac{3}{8\rho}$ is the standard BH entropy, $\tilde{S} = \frac{3}{8\rho f(\rho)}$ is modified BH entropy and $S_g$ is the entropy function of any entropic model. By specifying the function $S_g(S)$ from table \ref{chart}, one can reconstruct $f(\rho)$ and thus directly connect thermodynamic entropy to cosmological dynamics \cite{Khodam_Mohammadi_2023}.

\begin{table*}[h]
	\begin{center}
		\begin{tabular}{|c|c|c|c|}
			\hline
			No. & Entropy Type & $S_g$ & $f(\rho)$ \\
			\hline
			1 & Tsallis & $S_0(\dfrac{S}{S_0})^\delta$ & $(\dfrac{8}{3}S_0\rho)^{\delta-1}$ \\
			\hline
			2 & R\'enyi & $\dfrac{1}{\alpha}\ln(1+\alpha S)$ & $\dfrac{3\alpha}{8\rho}\left[\ln(1+\dfrac{3\alpha}{8\rho})\right]^{-1}$ \\
			\hline
			3 & Kaniadakis & $\dfrac{1}{K}\sinh(K S)$ & $\dfrac{3K}{8\rho}\left[\sinh(\dfrac{3K}{8\rho})\right]^{-1}$ \\
			\hline
		\end{tabular}
	\end{center}
	\caption{Correction functions $f(\rho)$ reconstructed from various entropy models \cite{Khodam_Mohammadi_2023}.} \label{chart}
\end{table*}

\section{Cosmography approach}\label{sec3}
Cosmographic analysis is a key tool for studying the dynamics of an FLRW universe in a model-independent manner. The cosmographic parameters appear in the Taylor expansion of the scale factor around the present time as follows \cite{Pourojaghi:2021den}:
\begin{eqnarray}\label{ceq2}
a(t) &\simeq & 1 + H_0 (t - t_0)\frac{q_0}{2!} H_0^2 (t - t_0)^2 + \frac{j_0}{3!} H_0^3 (t - t_0)^3 \nonumber \\
  &+& \frac{s_0}{4!} H_0^4 (t - t_0)^4 + \frac{l_0}{5!} H_0^5 (t - t_0)^5 + \frac{m_0}{6!} H_0^6 (t - t_0)^6 + \ldots
  \end{eqnarray}
  Here, the subscript `0' denotes the present-time values of the cosmographic parameters: deceleration ($q_0$), jerk ($j_0$), snap ($s_0$), lerk ($l_0$), and higher derivatives. The truncation error of the Taylor series in Eq. (\ref{ceq2}) affects the reconstruction accuracy of the universe's expansion model.

Furthermore, the cosmographic parameters can be expressed in terms of the Hubble parameter and its time derivatives as \cite{Visser:2003vq}:
\begin{eqnarray}\label{ceq3}
H^{(1)} &=& -H^2 (q + 1), \nonumber\\
H^{(2)} &=& H^3 (j + 3q + 2), \nonumber\\
H^{(3)} &=& H^4 \left( -4j - 3q^2 - 12q + s - 6 \right), \nonumber\\
H^{(4)} &=& H^5 \left( 10jq + 20j + l + 30q^2 + 60q - 5s + 24 \right),
\end{eqnarray}
where the superscript `$(n)$' denotes the $n$-th order time derivative of $H$.

The Taylor expansion of the dimensionless Hubble parameter, $E = H/H_0$, with respect to redshift $z = 1/a - 1$, can be written as \cite{Pourojaghi:2022zrh}:
\begin{eqnarray}\label{ceq4}
E(z) \simeq 1 + C_1 z + \frac{C_2 z^2}{2!} + \frac{C_3 z^3}{3!} + \frac{C_4 z^4}{4!} + \ldots,
\end{eqnarray}
where the coefficients $C_i$ are obtained from Eq. (\ref{ceq3}) as:
\begin{eqnarray}\label{ceq5}
C_1 &=& 1 + q_0, \nonumber\\
C_2 &=& j_0 - q_0^2, \nonumber\\
C_3 &=& -4j_0 q_0 - 3j_0 + 3q_0^3 + 3q_0^2 - s_0, \nonumber\\
C_4 &=& -4j_0^2 + 25j_0 q_0^2 + 32j_0 q_0 + 12j_0 + l_0 - 15q_0^4 - 24q_0^3 - 12q_0^2 + 7q_0 s_0 + 8s_0.
\end{eqnarray}

It is important to note that the Taylor expansion introduces significant errors, especially at higher redshifts ($z \gtrsim 1$), limiting its applicability in precision cosmology.

To address this, in modern cosmology, the Pad\'e approximation has emerged as a powerful tool for modeling cosmological observables such as the Hubble parameter, luminosity distance, and deceleration parameter. While Taylor series expansions are widely used, they suffer from poor convergence at higher redshifts and can lead to biased estimates of cosmographic parameters. In contrast, the Pad\'e approximation, formulated as a ratio of polynomials, offers significantly improved convergence properties and more accurate reconstructions of the cosmic expansion history over a wider redshift range.

A general Pad\'e approximant $P_{m,n}(z)$ is defined as \cite{Pourojaghi:2022zrh}:
\begin{equation}
P_{m,n}(z) = \frac{a_0 + a_1 z + \dots + a_m z^m}{1 + b_1 z + \dots + b_n z^n},
\end{equation}
where the coefficients are chosen to match the Taylor series of the function up to order $m + n$.

For example, for Pad\'e ($2,2$), we have \cite{Pourojaghi:2022zrh}
\begin{eqnarray}\label{eq9}
E(z)_{P_{2,2}} &=& \frac{a_0 + a_1 z + a_2 z^2}{1 + b_1 z + b_2 z^2}, \nonumber\\
a_0 &=& 1, \nonumber\\
a_1 &=& C_{1} + b_1, \nonumber\\
a_2 &=& C_{1} b_1 + \frac{C_{2}}{2} + b_2, \nonumber\\
b_1 &=& \frac{-C_{1}C_{4} + 2 C_{2} C_{3}}{4 C_{1} C_{3} - 6 C_{2}^2}, \nonumber\\
b_2 &=& \frac{3 C_{2} C_{4} - 4 C_{3}^2}{24 C_{1} C_{3} - 36 C_{2}^2}.
\end{eqnarray}
Here, $a_i$ and $b_i$ are coefficients corresponding to the Taylor expansion coefficients in Eq. (\ref{ceq5}). Two relations, Eqs. (\ref{ceq4}) and (\ref{eq9}), emphasize that by computing cosmographic quantities, one can approximately study the behavior of the function $E(z)$. However, the exact expression for $E(z)$ can be obtained directly from the Friedmann equation as follows. 

In a standard $\Lambda$CDM FRW cosmology without any radiation term, the dimensionless Hubble parameter is given by:
\begin{eqnarray}\label{eq11}
E(z)^{(\Lambda)} = \sqrt{ \Omega_{m0}(1+z)^3 + (1 - \Omega_{m0}) },
\end{eqnarray}
and from Eq. (\ref{ceq3}), the first four cosmographic parameters are calculated as:
\begin{eqnarray}\label{eq12}
q_0^{(\Lambda)} &=& -1 + \frac{3}{2} \Omega_{m0}, \nonumber\\
j_0^{(\Lambda)} &=& 1, \nonumber\\
s_0^{(\Lambda)} &=& 1 - \frac{9}{2} \Omega_{m0}, \nonumber\\
l_0^{(\Lambda)} &=& 1 + 3 \Omega_{m0} + \frac{27}{2} \Omega_{m0}^2.
\end{eqnarray}

For Tsallis cosmology in the old approach, we have \cite{Sheykhi:2018dpn}:
\begin{eqnarray}\label{eq13}
E(z)^{(T-O)} = \left[ \Omega_{m0}(1+z)^3 + (1 - \Omega_{m0}) \right]^{\frac{1}{4-2\delta}},
\end{eqnarray}
and the corresponding cosmographic parameters at present are given by \cite{Shahhoseini:2025pyr}:
\begin{eqnarray}\label{eq14}
q_0^{(T-O)} &=& \frac{3 \Omega_{m0}}{4 - 2\delta} - 1, \nonumber\\
j_0^{(T-O)} &=& 1 + \frac{9 (-1 + \delta) \Omega_{m0}^2}{2 (-2 + \delta)^2}, \nonumber\\
s_0^{(T-O)} &=& 1 + \frac{27 (-1 + \delta)(4\delta - 5) \Omega_{m0}^3}{4 (-2 + \delta)^3} - \frac{45 (-1 + \delta) \Omega_{m0}^2}{2 (-2 + \delta)^2} + \frac{9 \Omega_{m0}}{-4 + 2\delta}, \nonumber\\
l_0^{(T-O)} &=& 1 + \frac{81 (4\delta - 5)(-1 + \delta)(3\delta - 4) \Omega_{m0}^4}{4 (-2 + \delta)^4} - \frac{27 (-1 + \delta)(13\delta - 15) \Omega_{m0}^3}{(-2 + \delta)^3} \nonumber\\
&& + \frac{9 (28\delta - 25) \Omega_{m0}^2}{2 (-2 + \delta)^2} - \frac{3 \Omega_{m0}}{-2 + \delta}.
\end{eqnarray}

In the following, we derive the cosmographic quantities for non-extensive entropy cosmologies using the new approach.

\section{Cosmography of non-extensive Cosmology: New Approach}\label{sec4}
\subsection{Tsallis cosmology}

From Table \ref{chart}, we have:
\begin{eqnarray}
f(\rho) = \left( \frac{8}{3} S_0 \rho \right)^{\delta - 1}. \label{e3}
\end{eqnarray}
Substituting into the first Friedmann equation (\ref{FR}), we obtain:
\begin{eqnarray}
H^2 = \frac{\pi}{S_0} \left( \frac{8}{3} S_0 \rho \right)^\delta = H_0^2 \left( \frac{\rho}{\rho_{c0}} \right)^\delta, \label{e4}
\end{eqnarray}
where we set $H_0^2 = \pi / S_0$ and $\rho_{c0} = 3H_0^2 / 8\pi$. From the conservation equation (\ref{consf}),
\begin{eqnarray}
\dot{\rho_i} \delta + 3H \rho_i (1 + w_i) = 0, \label{e5}
\end{eqnarray}
which integrates to
\begin{eqnarray}
\rho_i = \rho_{i0} (1 + z)^{\frac{3}{\delta}(1 + w_i)}. \label{e6}
\end{eqnarray}

Using Eq. (\ref{e6}) in Eq. (\ref{e4}), for a cosmic fluid contain dark matter (DM) and vacuum energy ($\Lambda$), $\Lambda$CDM ($w_d = -1$) without radiation, we find:
\begin{eqnarray}
E(z)^{(T-N)} = \frac{H}{H_0} = \left[ \Omega_{m0} (1+z)^{\frac{3}{\delta}} + (1 - \Omega_{m0}) \right]^{\delta/2}. \label{e7}
\end{eqnarray}

Using Eq. (\ref{ceq3}), the cosmographic parameters in this new approach are then calculated as:
\begin{eqnarray}
q_0^{(T-N)} &=& -1 + \frac{3}{2} \Omega_{m0}, \nonumber\\
j_0^{(T-N)} &=& 1 + \frac{9 \Omega_{m0}}{2\delta} (1 - \Omega_{m0})(1 - \delta), \nonumber\\
s_0^{(T-N)} &=& 1 + \frac{9 \Omega_{m0}}{4 \delta^2} \Big[ (-9 \Omega_{m0}^2 + 11 \Omega_{m0} - 4) \delta^2 - (21 \Omega_{m0} - 8)(1 - \Omega_{m0}) \delta \nonumber\\
&& + (2 \Omega_{m0} - 1)(1 - \Omega_{m0}) \Big], \nonumber\\
l_0^{(T-N)} &=& 1 + \frac{81 \Omega_{m0}}{2 \delta^3} \Big[ \frac{1}{2} \Omega_{m0}^3 (\delta - 1)(3\delta - 4)(2\delta - 3) - \frac{1}{6} \Omega_{m0}^2 (\delta - 1)(25\delta^2 - 89\delta + 72) \nonumber\\
&& + \Omega_{m0} \left( \frac{35}{18} \delta^3 - \frac{155}{18} \delta^2 + 14\delta - 7 \right) - \frac{10}{27} \delta^3 + \frac{10}{9} \delta^2 - \frac{5}{3} \delta + 1 \Big]. \label{e11}
\end{eqnarray}

It is worth mentioning that in both approaches, Eqs. (\ref{e11}) and (\ref{eq14}), all cosmographic quantities and the dimensionless Hubble function merged to their corresponding standard cosmology values, Eq. (\ref{eq12}), in the limiting case, $\delta = 1$.

\subsection{R\'enyi Cosmology}
From Table \ref{chart},
\begin{eqnarray}
f(\rho) = \frac{3\alpha}{8\rho} \left[ \ln\left(1 + \frac{3\alpha}{8\rho} \right) \right]^{-1}. \label{e12}
\end{eqnarray}
The first Friedmann equation (\ref{FR}) then becomes:
\begin{eqnarray}
H^2 = \alpha\pi \left[ \ln\left(1 + \frac{3\alpha}{8\rho} \right) \right]^{-1} = H_0^2 \left[ \ln\left(1 + \frac{\rho_{c0}}{\rho} \right) \right]^{-1}, \label{e13}
\end{eqnarray}
where we set $H_0^2 = \alpha\pi$ and $\rho_{c0} = 3H_0^2 / 8\pi$. Inserting Eq. (\ref{e12}) into the continuity equation (\ref{consf}) yields:
\begin{eqnarray}
\frac{\dot{u}_i}{u_i \ln(u_i)} - 3H(1 + w_i) = 0, \label{e14}
\end{eqnarray}
where $u_i = 1 + \rho_{c0}/\rho_i = 1 + 1/v_i$, with $v_i = \rho_i / \rho_{c0}$. Solving this equation gives:
\begin{eqnarray}
v_i = \frac{1}{ \exp\left[ A_{i0} (1 + z)^{-3(1 + w_i)} \right] - 1 }. \label{e15}
\end{eqnarray}

Substituting Eq.~~(\ref{e15}) into Eq. (\ref{e13}), the dimensionless Hubble parameter for matter-vacuum ($w_m=0, ~w_d=-1$) becomes:
\begin{eqnarray}
E = \left\{ \ln\left[ 1 + \frac{1}{v_m + v_d} \right] \right\}^{-1/2},\label{e16}
\end{eqnarray}
where 
\begin{eqnarray}
v_m &=& \left\{ \exp\left[ A_{m0} (1 + z)^{-3} \right] - 1 \right\}^{-1}, \nonumber\\
v_d &=& \left\{ \exp(A_{d0}) - 1 \right\}^{-1}. \label{e17}
\end{eqnarray}

The density parameters are related via:
\begin{eqnarray}
\Omega_{m0} &=& \frac{\rho_{m0}}{\rho_{c0}} = \left[ \exp(A_{m0}) - 1 \right]^{-1}, \nonumber\\
\Omega_{d0} &=& \left[ \exp(A_{d0}) - 1 \right]^{-1},
\end{eqnarray}
where $A_{d0}$ is determined by imposing $E = 1$ at $z = 0$ in Eq.\~(\ref{e16}), yielding:
\begin{equation}
A_{d0} = \ln\left[ \frac{ \exp(A_{m0} + 1) - 2e + 1 }{ \exp(A_{m0}) - e } \right].
\end{equation}

Thus, Eq.\~(\ref{e16}) can be rewritten as:
\begin{eqnarray}
E(z)^{(R)} &=& \left\{ \ln\left[ \frac{ e^{A_{m0}(1 + z)^{-3}}(2e - e^{A_{m0}+1} - 1) + e^{A_{m0}} - e }{ e^{A_{m0}(1 + z)^{-3}}(e - e^{A_{m0}}) + 2e^{A_{m0}} - e^{A_{m0}+1} - 1 } \right] \right\}^{-1/2}\notag\\
A_{m0}&=&\ln(1+\frac{1}{\Omega_{m0}}). \label{e166}
\end{eqnarray}

It is noteworthy that $E(z)$ is expressed solely in terms of the parameter $A_{m0}$, which itself is a function of $\Omega_{m0}$.

The present-time cosmographic quantities are obtained as:
\begin{eqnarray}
q_0^{(R)} &=& -1 + \frac{3 A_{m0} e^{A_{m0}-1} (e - 1)^2 }{ 2 (e^{A_{m0}} - 1)^2 }, \nonumber\\
j_0^{(R)} &=& 1 + \frac{9 A_{m0} e^{A_{m0}-1} (e - 1)^2 }{ 2 (e^{A_{m0}} - 1)^4 } \Big[ (e^{2A_{m0}} + e^{A_{m0}+1} - 4e^{A_{m0}} + 3e^{A_{m0}-1} - 1)A_{m0} \nonumber\\
&& - 2e^{2A_{m0}} + 4e^{A_{m0}} - 2 \Big].
\end{eqnarray}

It should be noted that the other cosmographic parameters $s_0^{(R)}$ and $l_0^{(R)}$ have been computed using Maple code but are too lengthy to include here. All cosmographic parameters depend on $A_{m0}$. The Hubble rate $E(z^{(R)})$ can also be approximated via the Pad\'e $(2,2)$ formula (\ref{eq9}), ultimately expressed in terms of $\Omega_{m0}$.

\subsection{Kaniadakis Cosmology}

From Table \ref{chart},
\begin{eqnarray}
f(\rho) = \frac{3k}{8\rho} \left[ \sinh\left( \frac{3k}{8\rho} \right) \right]^{-1}. \label{e20}
\end{eqnarray}
The first Friedmann equation (\ref{FR}) then becomes:
\begin{eqnarray}
H^2 = k\pi \left[ \sinh\left( \frac{3k}{8\rho} \right) \right]^{-1} = H_0^2 \left[ \sinh\left( \frac{\rho_{c0}}{\rho} \right) \right]^{-1}, \label{e21}
\end{eqnarray}
where $H_0^2 = k\pi$ and $\rho_{c0} = 3H_0^2 / 8\pi$. Inserting Eq.\~(\ref{e20}) into the continuity equation (\ref{consf}) yields:
\begin{eqnarray}
\dot{u}_i \coth(u_i) - 3H(1 + w_i) = 0, \label{e22}
\end{eqnarray}
where $u_i = \rho_{c0}/\rho_i = 1/v_i$, with $v_i = \rho_i / \rho_{c0}$. Solving this equation gives:
\begin{eqnarray}
v_i = \left[ \text{arcsinh}\left( B_{i0}(1 + z)^{-3(1 + w_i)} \right) - 1 \right]^{-1}. \label{e23}
\end{eqnarray}

The dimensionless Hubble parameter for matter-vacuum ($w_m=0, ~w_d=-1$) is:
\begin{eqnarray}
E = \left[ \sinh\left( \frac{1}{v_m + v_d} \right) \right]^{-1/2}, \label{e24}
\end{eqnarray}
where
\begin{eqnarray}
v_m &=& \left[ \text{arcsinh}\left( B_{m0}(1 + z)^{-3} \right) \right]^{-1}, \nonumber\\
v_d &=& \left[ \text{arcsinh}(B_{d0}) \right]^{-1}. \label{e25}
\end{eqnarray}

The density parameters are related via:
\begin{eqnarray}
\Omega_{m0} &=& \frac{\rho_{m0}}{\rho_{c0}} = \left[ \text{arcsinh}(B_{m0}) \right]^{-1}, \nonumber\\
\Omega_{d0} &=& \left[ \text{arcsinh}(B_{d0}) \right]^{-1},
\end{eqnarray}
with $B_{d0}$ determined by setting $E = 1$ at $z = 0$ in Eq.\~(\ref{e24}), giving:
\begin{eqnarray}
B_{d0} = \sinh\left\{ \left[ \left( \ln(1+\sqrt{2}) \right)^{-1} - \left( \text{arcsinh}(B_{m0}) \right)^{-1} \right]^{-1} \right\}.
\end{eqnarray}

Thus, $E$ can be written entirely in terms of the single variable $B_{m0}$, which depends on $\Omega_{m0}$:
\begin{eqnarray}
E(z)^{(K)} &=& \left\{ \sinh\left[ \frac{1}{ \text{arcsinh}(B_{m0}(1+z)^{-3}) } - \frac{1}{ \text{arcsinh}(B_{m0}) } + \frac{1}{\ln(1+\sqrt{2})} \right]^{-1} \right\}^{-1/2}, \notag\\
B_{m0}&=& \sinh(\frac{1}{\Omega_{m0}}) \label{e244}
\end{eqnarray}

The present-time cosmographic parameters, assuming vacuum dark energy ($w_d = -1$), are:
\begin{eqnarray}
q_0^{(K)} &=& -1 + \frac{3(2 + \sqrt{2}) B_{m0} \ln^2(1 + \sqrt{2}) }{ \text{arcsinh}^{2}(B_{m0}) \sqrt{B_{m0}^2 + 1} (2 + 2\sqrt{2}) }, \nonumber\\
j_0^{(K)} &=& 1 + \frac{27 B_{m0} \ln^2(1 + \sqrt{2}) }{ \text{arcsinh}^{4}(B_{m0}) (B_{m0}^2 + 1)^2 (4\sqrt{2} + 6) } \Big\{
2 B_{m0}(B_{m0}^2 + 1) \Big[ \left( \sqrt{2} + \frac{3}{2} \right) \ln^2(1 + \sqrt{2}) \nonumber\\
&& - \left( \sqrt{2} + \frac{4}{3} \right) \ln(1 + \sqrt{2}) + \text{arcsinh}(B_{m0}) \left( \sqrt{2} + \frac{4}{3} \right) \Big] \nonumber\\
&& - (B_{m0}^2 + 2)\left( \sqrt{2} + \frac{4}{3} \right) \text{arcsinh}^{2}(B_{m0}) \sqrt{B_{m0}^2 + 1} \Big\}.
\end{eqnarray}

Same as R\'enyi cosmology, the other cosmographic parameters $s_0^{(K)}$ and $l_0^{(K)}$ have been computed using Maple code but are too lengthy to include here. All cosmographic parameters depend on $B_{m0}$. Also the Hubble rate $E(z)^{(K)}$ can be approximated via the Pad\'e $(2,2)$ formula (\ref{eq9}), ultimately expressed in terms of $\Omega_{m0}$.

\section{Physical meaning of the matter-sector conformal rescaling}

In this section we explain what we mean by ``conformal transformation on the right-hand side (RHS)'' of Einstein equations, and why it is needed when we use non-extensive entropies in cosmology. Initially, by comparing the standard Friedmann equation with the modified Friedmann equation that has been derived directly from non-extensive entropy and the laws of thermodynamics at the apparent horizon (see Eq. (\ref{fr-thermo})), we made a natural conjecture \cite{Khodam-Mohammadi:2023PLB}. We suggested that these changes may come from a conformal rescaling of the right-hand side of Einstein's equations, that is, the energy–momentum tensor. This idea was already mentioned in our previous work \cite{Khodam-Mohammadi:2023PLB}. The important question, however, is whether this mapping is just made by hand or if it has physical support. In the following, we present arguments that show this mapping has real physical justification and is not only an assumption.

First, from the point of view of conformal rescaling in the matter sector, in our model we do not change the geometry side of Einstein equations. We start from
\begin{equation}
G_{\mu\nu} = 8\pi G\,T_{\mu\nu},
\end{equation}
and we introduce an effective matter source
\begin{equation}
T^{\text{(eff)}}_{\mu\nu} = f(\rho)\,T_{\mu\nu}, \qquad f(\rho)>0,
\end{equation}
where the function $f(\rho)$ is fixed by the chosen non-extensive entropy through horizon thermodynamics.  

For example, in the Tsallis case (see Table \ref{chart}), we have
\begin{equation}
f(\rho) = \left(\tfrac{8}{3}S_0\,\rho\right)^{\delta-1}.
\end{equation}
This rescaling can be seen as a Weyl (conformal) rescaling but only in the matter sector. This fact can be clearly examined in scalar-tensor theories, which will be presented later.

Second, from the thermodynamic derivation point of view, we start from the Clausius relation at the apparent horizon,
\begin{equation}
\delta Q = T\,dS.
\end{equation}
When entropy $S$ is non-extensive (for example Tsallis $S\sim A^\delta$), the heat flow $\delta Q$ corresponds to the energy flux of the effective matter source. Compatibility with the Bianchi identity requires
\begin{equation}
\nabla^\mu T^{\text{(eff)}}_{\mu\nu}=0,
\end{equation}
which leads to the modified continuity equation
\begin{equation}
\dot{\rho}\,\delta + 3H\rho(1+w)=0.
\end{equation}
The solution is
\begin{equation}
\rho(a)=\rho_{0}\,a^{-\tfrac{3}{\delta}(1+w)},
\end{equation}
which is given in Eq.(\ref{e6}).  

From this, the first Friedmann equation follows,
\begin{equation}
H^2 = \frac{\pi}{S_0}\Big(\tfrac{8}{3}S_0\rho\Big)^\delta 
     = H_0^2 \Big(\tfrac{\rho}{\rho_{c0}}\Big)^\delta, 
\end{equation}
and for $\Lambda$CDM (no radiation), the dimensionless Hubble parameter is
\begin{equation}
E(z) = \Big[ \Omega_{m0}(1+z)^{\tfrac{3}{\delta}} + (1-\Omega_{m0}) \Big]^{\delta/2}.
\end{equation}
So the function $f(\rho)$ and the modified dynamics are not free assumptions. They come directly from the non-extensive entropy law. In the limit $\delta=1$ we recover general relativity:
\begin{equation}
f(\rho)\to 1, \quad 
\rho \propto a^{-3(1+w)}, \quad 
E(z)\to \sqrt{\Omega_{m0}(1+z)^3 + (1-\Omega_{m0})}.
\end{equation}

Third, in relation to other frameworks, we note that this conformal mapping is analogous to the Einstein-frame rescaling in scalar-tensor gravity \cite{Damour:1996ke,Fujii:2003pa,Faraoni:2004pi}, where $\tilde g_{\mu\nu}=A^2(\phi)g_{\mu\nu}$ and the stress tensor is rescaled. Our case is the same but with $A$ replaced by a density-dependent factor fixed by entropy. It also resembles nonequilibrium or particle-creation cosmologies where the conservation law is modified \cite{Prigogine:1989zz,Lima:1995xz,Pan:2017zoh}, which in our model appears as $\delta\neq 1$. It may also be interpreted as an effective running coupling or fluid redefinition \cite{Sola:2013gha,Wetterich:2013jsa}, but here it is uniquely fixed by the choice of non-extensive entropy \cite{Cai:2005ra,Moradpour:2018ivi}.

Finally, about energy conditions and stability, for $\delta>0$, $f(\rho)>0$ for $\rho>0$, so there is no ghost problem. The effective density and pressure are
\begin{equation}
\rho_{\rm eff}=f\rho,\qquad p_{\rm eff}=f p,
\end{equation}
so the null energy condition holds.  

For a barotropic fluid, $p=w\rho$, the effective sound speed for a constant $w$ is
\begin{equation}
c_{s,\rm eff}^2 = \frac{dp_{\rm eff}}{d\rho_{\rm eff}} = w,
\end{equation}
which is the same as the standard fluid. Therefore, this mapping does not impose any new instability on the cosmological system.

In summary, the conformal rescaling of the matter sector not only is  an arbitrary assumption but also is the direct consequence of applying non-extensive entropy thermodynamics to the apparent horizon. In this picture, the gravitational side of Einstein equations is unchanged, but the matter side is modified in a self-consistent way. This is different from ad-hoc modifications of gravity (for example, $f(R)$ or higher-order terms), where the geometry side of Einstein equations is changed by hand. At the limit where entropy reduces to the standard Bekenstein-Hawking form (for Tsallis $\delta=1$), we have $f(\rho)=1$, and the effective tensor becomes the usual $T_{\mu\nu}$ where the Einstein gravity with standard cosmology is fully recovered. It should be also emphasized that in this framework the microscopic density of the cosmic fluid remains $\rho=m~n$. What changes is the effective energy density that couples to geometry, given by $\rho_{eff}=f(\rho)\rho$. This shows that the influence of non-extensive entropy is not a redefinition of matter itself but a modification of its gravitational imprint, which vanishes in the limit $f(\rho)=1$. This shows that our approach is consistent with general relativity but extends it naturally to non-extensive entropic cases.

\section{Results Analysis}\label{sec5}
As mentioned, in both approaches, the dimensionless Hubble function $E(z)$ and cosmographic parameters can be calculated within the framework of non-extensive entropy cosmology in the new approach. An important point is that in the limit $\delta = 1$, both the new and old approaches merge to standard cosmology.

This leads to a fundamental question: which approach is more appropriate for describing the universe's expansion? In particular, how do the key cosmographic parameters behave as functions of redshift $z$? At which redshift ranges does the tension between the two approaches become most significant?

\begin{figure*} [!h]
\begin{center}
\includegraphics[width=8cm]{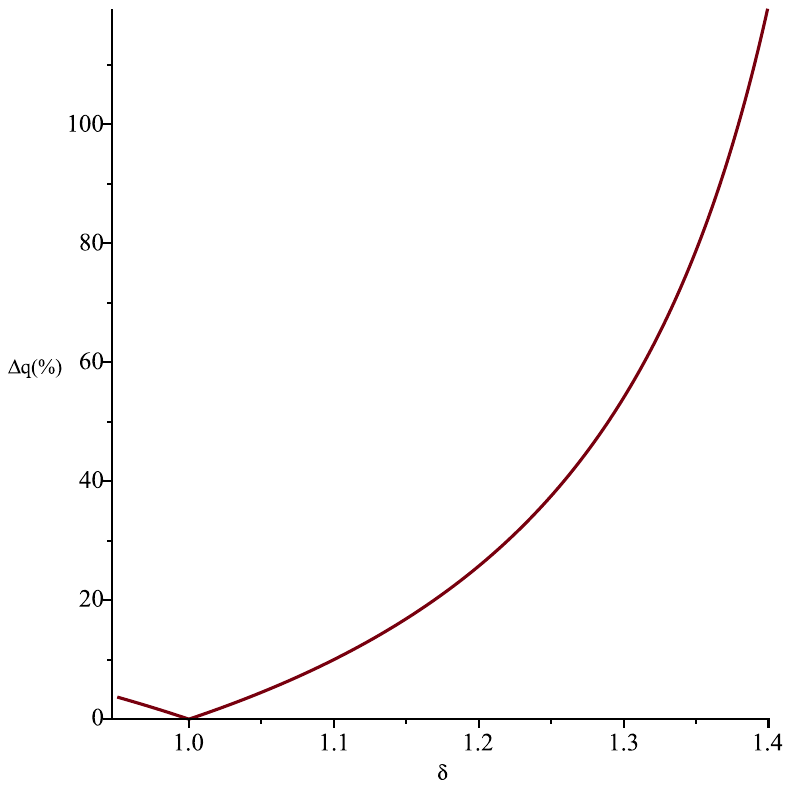}
\includegraphics[width=8cm]{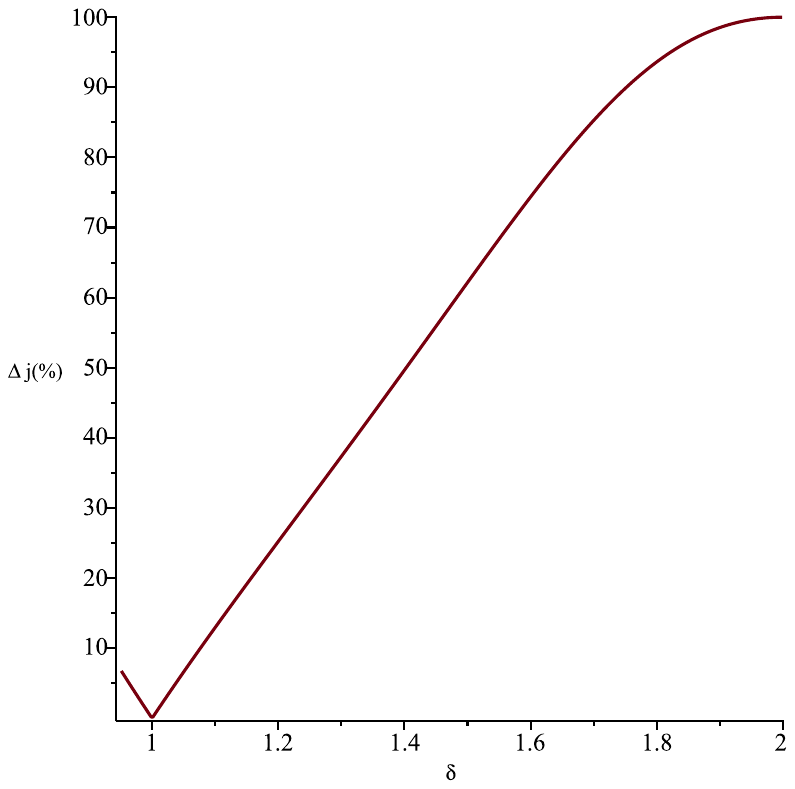}
\includegraphics[width=8cm]{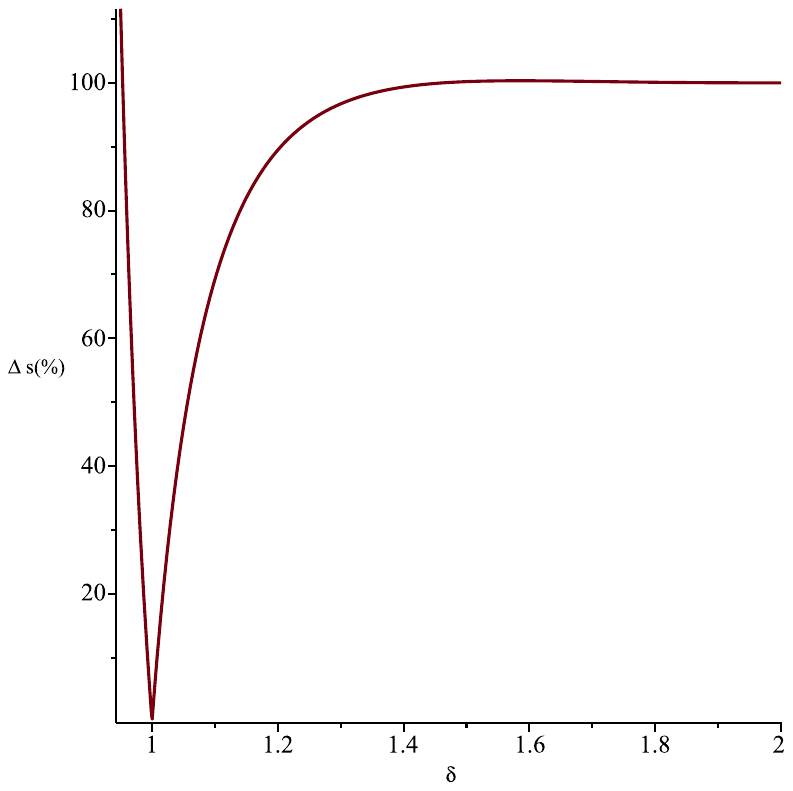}
\includegraphics[width=8cm]{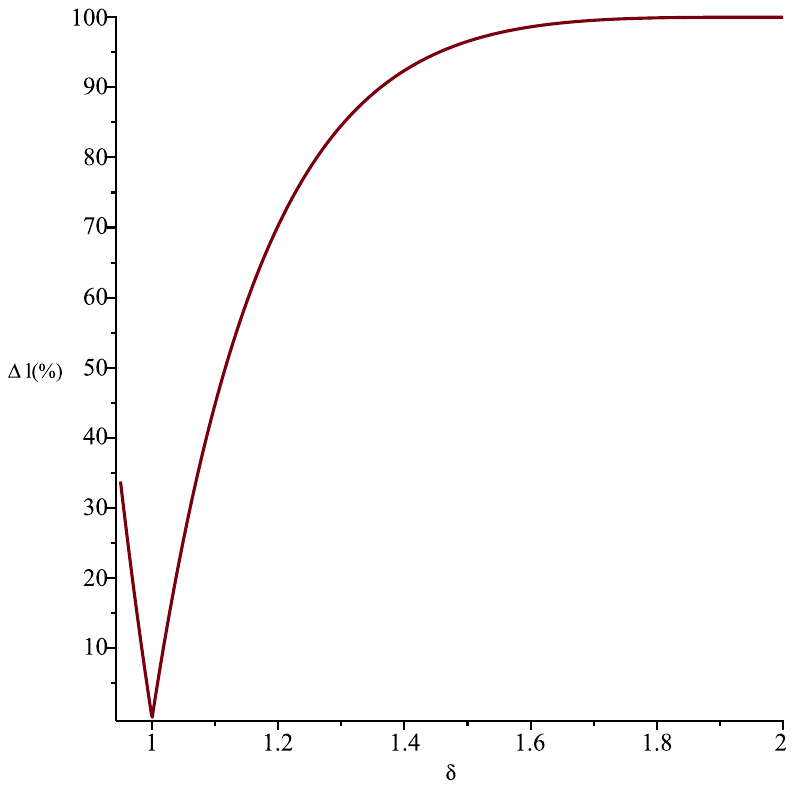}
\vspace{-100pt}
\caption{Relative percentage difference of cosmographic parameters versus $\delta$ for Tsallis cosmology with $\Omega_{m0}=0.3$, comparing the new approach, old approach, and the $\Lambda$CDM model.}
\label{fig1}
\end{center}
\end{figure*}

\begin{figure}[!h]
\centering
\includegraphics[scale=0.6]{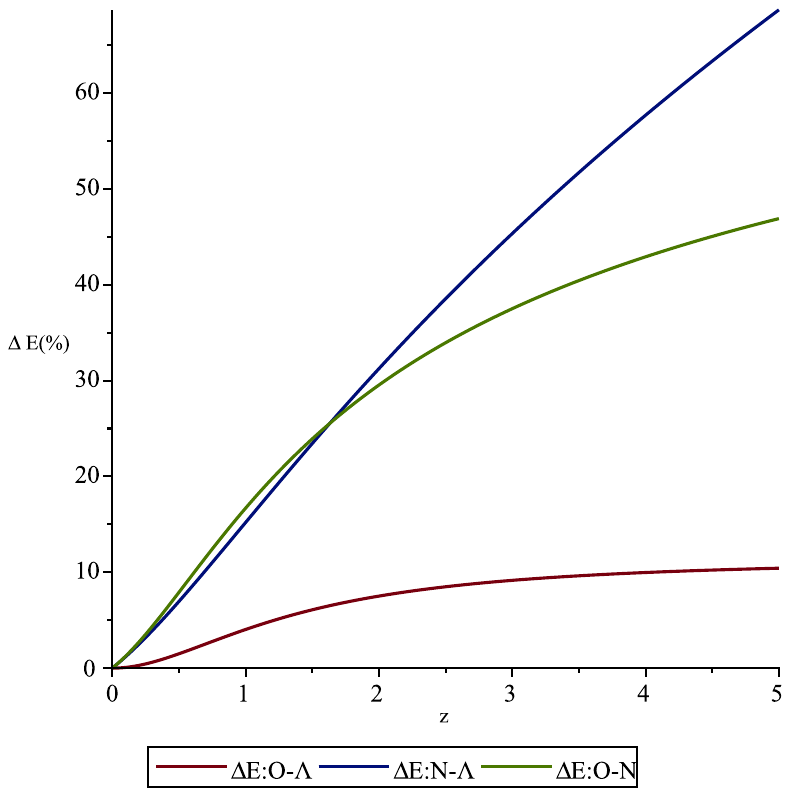}
\vspace{-200pt}
\caption{Relative percentage difference of the dimensionless Hubble parameter $E(z)$ versus redshift $z$ for Tsallis cosmology with $\Omega_{m0} = 0.3$ and $\delta=1.2$, comparing the new approach, old approach, and the $\Lambda$CDM model.}
\label{fig2}
\end{figure}
To address these questions, we examined the relative percentage difference ($\Delta \Gamma$) of all cosmographic parameters ($\Gamma: q_0,~j_0,~s_0,~l_0$) calculated from both approaches in the Tsallis model, as a function of different $\delta$ values. We also analyzed the relative percentage difference for the Hubble function $E(z)$ between the two approaches and standard FRW cosmology, as a function of redshift $z$.

In Fig. \ref{fig1}, the behavior of the relative percentage difference ($\Delta \Gamma$) for each cosmographic parameter is shown, defined by:
\begin{equation}
\Delta \Gamma = \left| \frac{\Gamma_O - \Gamma_N}{\Gamma_O} \right| \times 100,
\end{equation}
where $\Gamma_O$ and $\Gamma_N$ denote values for the old and new approaches, respectively.

As seen in this figure, at $\delta = 1$, the differences in all cosmographic parameters between the two approaches vanish ($\Delta = 0$). The value of $\Delta$ increases on both sides of $\delta = 1$, where it reaching to $\Delta = 100\% $ at right, rapidly for Tsallis parameter values $\delta < 2 $. Additionally, in the top-left plot of Fig. \ref{fig1}, a singularity appears at $\delta \simeq 1.5$, indicating that the most suitable range for the Tsallis parameter is $\delta < 1.5 $ to avoid divergences.

In Fig. \ref{fig2}, we analyze $\Delta E^{(T)}$ for three model comparisons: $O-\Lambda$ (old approach Tsallis Vs. standard FRW, red line), $N-\Lambda$ (new approach Tsallis Vs. standard FRW, green line), and $O-N$ (old approach Tsallis Vs. new approach Tsallis, blue line), plotted against redshift $z$ for $\Omega_{m0}=0.3$ and $\delta=1.2$. As shown, the tension in $E(z)^{(T)}$ relative to standard FRW increases in the new approach across all $z$ values. Furthermore, $\Delta$ grows with increasing $z$, and the tension between the old and new approaches is particularly large at high redshifts. Therefore, while at small $z$ all approaches yield similar results, the differences become significant at higher redshifts.

\begin{figure}[!h]
\centering
\includegraphics[scale=0.6]{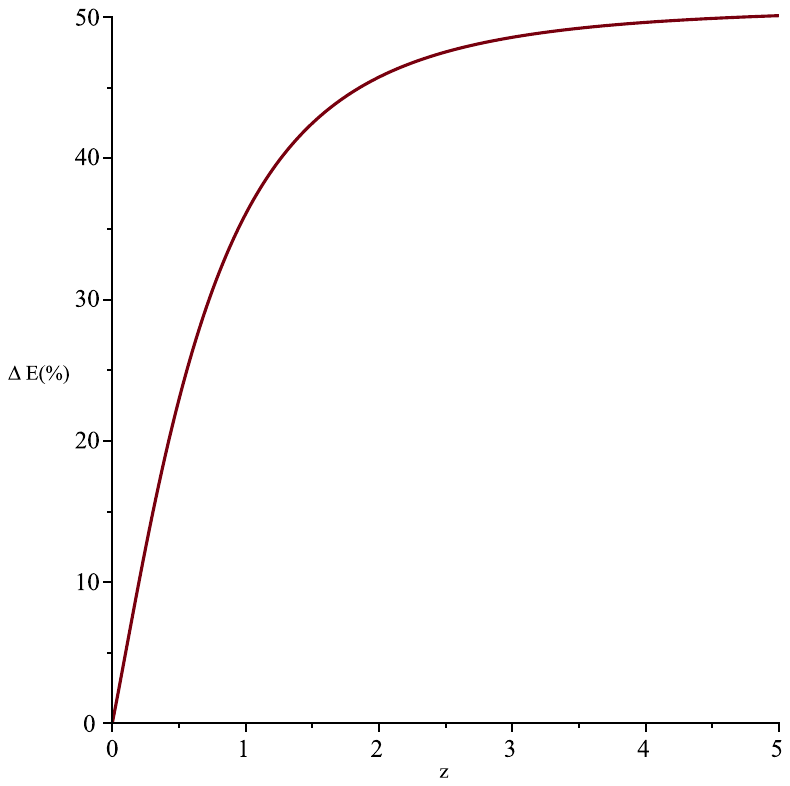}
\vspace{-200pt}
\caption{Relative percentage difference of the dimensionless Hubble parameter $E(z)$ versus redshift $z$ for R\'enyi cosmology with $\Omega_{m0} = 0.3$, compared to the $\Lambda$CDM model.}
\label{fig3}
\end{figure}

\begin{figure}[!h]
\centering
\includegraphics[scale=0.6]{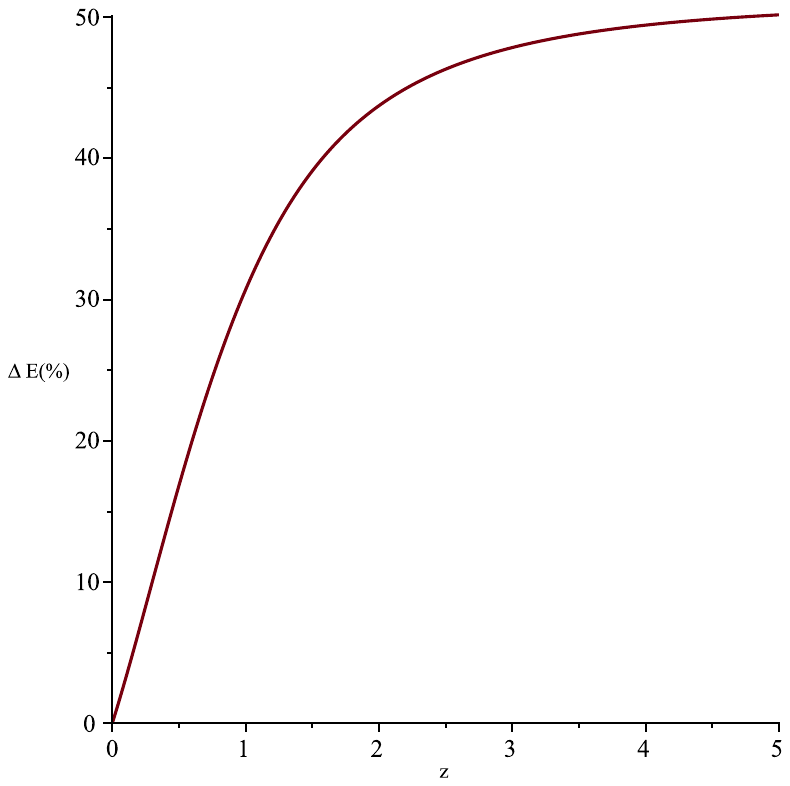}
\vspace{-200pt}
\caption{Relative percentage difference of the dimensionless Hubble parameter $E(z)$ versus redshift $z$ for Kaniadakis cosmology with $\Omega_{m0} = 0.3$, compared to the $\Lambda$CDM model.}
\label{fig4}
\end{figure}

The tension between R\'enyi cosmology and standard FRW cosmology is illustrated in Fig. \ref{fig3} for $\Omega_{m0}=0.3$. As shown, $\Delta E^{(R-\Lambda)}$ increases with redshift $z$: it remains small at low redshift but exhibits significant tension at high redshift. A similar behavior is observed in Kaniadakis cosmology in Fig. \ref{fig4}. An important point about these two entropic models is that they each contain only a single free parameter, $\Omega_{m0}$, similar to the $\Lambda$CDM model. 

These differences in cosmographic parameters and expansion rates indicate that the modified approach can alter the inferred Hubble constant compared to $\Lambda$CDM. This provides a framework that could potentially alleviate or resolve the Hubble tension by modifying the effective expansion rate at low and intermediate redshifts in a manner consistent with observations. 

\section{Conclusion}
In this work, we conducted a comprehensive cosmographic analysis of Friedmann cosmologies modified by non-extensive entropic frameworks, including Tsallis, R\'enyi, and Kaniadakis entropies, within a novel modified energy-momentum tensor approach. Also we clarified the physical meaning of the conformal rescaling in the matter sector. This mapping is not an arbitrary modification but follows directly from non-extensive entropy thermodynamics, ensuring consistency with general relativity while extending it to more non-extensive entropic frameworks. We stress that our model does not alter the microscopic definition of matter density $\rho=m~n$, instead the non-extensive entropy at the apparent horizon modifies the effective source in the Friedmann equation to $\rho_{eff}=f(\rho)\rho$. Thus, the universe evolves as if it were driven by $\rho_{eff}$, while the standard definition of matter remains unchanged. By deriving generalized Friedmann equations for each entropy model, we obtained analytical expressions for key cosmographic parameters such as the deceleration ($q_0$), jerk ($j_0$), snap ($s_0$), and lerk ($l_0$), and examined their behavior relative to the standard $\Lambda$CDM model.

Our analysis revealed significant differences between the traditional formal approach and the modified energy-momentum approach in Tsallis cosmology. The relative percentage differences in cosmographic parameters increase sharply for deformation parameter values deviating from unity, with a singularity appearing around $\delta \simeq 1.5$, restricting the viable parameter space to $\delta < 1.5$ to avoid singularity on $\Delta~E$. Additionally, the relative differences in the dimensionless Hubble function $E(z)$ between the two approaches and $\Lambda$CDM grow with increasing redshift, indicating that observational data at high redshifts can effectively discriminate between these formulations. The difference between old and new approach appears in higher cosmographic parameters ($j_0, s_0, l_0$), see Eqs.(\ref{eq14}) and (\ref{e11}). Our plots show that relative difference $\Delta \Gamma$ and $\Delta E$ grow with $|\delta-1|$ and with redshift. This means the two approaches are observationally distinguishable, and not only a formal change.

In the cases of R\'enyi and Kaniadakis cosmologies, both models exhibit expansion behavior closely matching $\Lambda$CDM at low redshifts but show increasing tension at higher redshifts. An important feature of these models is that they each have only a single free parameter, $\Omega_{m0}$, similar to $\Lambda$CDM, enhancing their simplicity and potential viability as alternative descriptions of cosmic acceleration.

Overall, the modified energy-momentum tensor approach extends applicability to a broader class of non-extensive entropy models, enabling consistent treatment of R\'enyi and Kaniadakis cosmologies that were previously intractable under the formal thermodynamic method. The significant deviations found in Tsallis cosmology between the two approaches underscore the importance of model selection and the capability of high-precision data to test these frameworks.

Given the persistent $\sim5\sigma$ Hubble tension between local Cepheid-calibrated measurements and global CMB-inferred values, our results suggest that modified entropic cosmologies could potentially alleviate this discrepancy by altering the effective expansion rate at low and intermediate redshifts while remaining consistent with early-universe constraints. This positions non-extensive entropic models as promising candidates for addressing fundamental tensions in precision cosmology within a thermodynamic-gravity framework.

Future research should focus on performing full likelihood analyses incorporating Type Ia supernovae (Pantheon+), baryon acoustic oscillations (DESI), and cosmic microwave background data (Planck) to place stringent constraints on deformation parameters and rigorously test the viability of these models against standard cosmology. Furthermore, investigating perturbation dynamics and structure formation within these frameworks will provide a comprehensive assessment of their cosmological consistency and potential to resolve current observational tensions.

\bibliography{refth}

\begin{thebibliography}{75}
\expandafter\ifx\csname natexlab\endcsname\relax\def\natexlab#1{#1}\fi
\expandafter\ifx\csname bibnamefont\endcsname\relax
  \def\bibnamefont#1{#1}\fi
\expandafter\ifx\csname bibfnamefont\endcsname\relax
  \def\bibfnamefont#1{#1}\fi
\expandafter\ifx\csname citenamefont\endcsname\relax
  \def\citenamefont#1{#1}\fi
\expandafter\ifx\csname url\endcsname\relax
  \def\url#1{\texttt{#1}}\fi
\expandafter\ifx\csname urlprefix\endcsname\relax\def\urlprefix{URL }\fi
\providecommand{\bibinfo}[2]{#2}
\providecommand{\eprint}[2][]{\url{#2}}

\bibitem[{\citenamefont{Hawking}(1975)}]{Hawking:1975vcx}
\bibinfo{author}{\bibfnamefont{S.~W.} \bibnamefont{Hawking}},
  \bibinfo{journal}{Commun. Math. Phys.} \textbf{\bibinfo{volume}{43}},
  \bibinfo{pages}{199} (\bibinfo{year}{1975}), \bibinfo{note}{[Erratum:
  Commun.Math.Phys. 46, 206 (1976)]}.

\bibitem[{\citenamefont{Bekenstein}(1973)}]{Bekenstein:1973ur}
\bibinfo{author}{\bibfnamefont{J.~D.} \bibnamefont{Bekenstein}},
  \bibinfo{journal}{Phys. Rev. D} \textbf{\bibinfo{volume}{7}},
  \bibinfo{pages}{2333} (\bibinfo{year}{1973}).

\bibitem[{\citenamefont{Jacobson}(1995)}]{Jacobson:1995ab}
\bibinfo{author}{\bibfnamefont{T.}~\bibnamefont{Jacobson}},
  \bibinfo{journal}{Phys. Rev. Lett.} \textbf{\bibinfo{volume}{75}},
  \bibinfo{pages}{1260} (\bibinfo{year}{1995}).

\bibitem[{\citenamefont{Shahhoseini
  et~al.}(2025{\natexlab{a}})\citenamefont{Shahhoseini, Malekjani, and
  Khodam-Mohammadi}}]{Shahhoseini:2025sgl}
\bibinfo{author}{\bibfnamefont{N.}~\bibnamefont{Shahhoseini}},
  \bibinfo{author}{\bibfnamefont{M.}~\bibnamefont{Malekjani}},
  \bibnamefont{and}
  \bibinfo{author}{\bibfnamefont{A.}~\bibnamefont{Khodam-Mohammadi}},
  \bibinfo{journal}{Eur. Phys. J. C} \textbf{\bibinfo{volume}{85}},
  \bibinfo{pages}{53} (\bibinfo{year}{2025}{\natexlab{a}}),
  \eprint{2501.03655}.

\bibitem[{\citenamefont{Brevik and Timoshkin}(2024)}]{Brevik:2024nzf}
\bibinfo{author}{\bibfnamefont{I.}~\bibnamefont{Brevik}} \bibnamefont{and}
  \bibinfo{author}{\bibfnamefont{A.~V.} \bibnamefont{Timoshkin}},
  \bibinfo{journal}{Int. J. Geom. Meth. Mod. Phys.}
  \textbf{\bibinfo{volume}{21}}, \bibinfo{pages}{2450181}
  (\bibinfo{year}{2024}), \eprint{2404.05597}.

\bibitem[{\citenamefont{Sheykhi and Ghaffari}(2023)}]{Sheykhi_2023}
\bibinfo{author}{\bibfnamefont{A.}~\bibnamefont{Sheykhi}} \bibnamefont{and}
  \bibinfo{author}{\bibfnamefont{S.}~\bibnamefont{Ghaffari}},
  \bibinfo{journal}{Physics of the Dark Universe}
  \textbf{\bibinfo{volume}{41}}, \bibinfo{pages}{101241}
  (\bibinfo{year}{2023}), ISSN \bibinfo{issn}{2212-6864},
  \urlprefix\url{http://dx.doi.org/10.1016/j.dark.2023.101241}.

\bibitem[{\citenamefont{P et~al.}(2023)\citenamefont{P, Pandey, Sharma, and
  Pankaj}}]{P:2022amn}
\bibinfo{author}{\bibfnamefont{S.~K.} \bibnamefont{P}},
  \bibinfo{author}{\bibfnamefont{B.~D.} \bibnamefont{Pandey}},
  \bibinfo{author}{\bibfnamefont{U.~K.} \bibnamefont{Sharma}},
  \bibnamefont{and} \bibinfo{author}{\bibnamefont{Pankaj}},
  \bibinfo{journal}{Eur. Phys. J. C} \textbf{\bibinfo{volume}{83}},
  \bibinfo{pages}{143} (\bibinfo{year}{2023}), \eprint{2211.15468}.

\bibitem[{\citenamefont{Nojiri et~al.}(2022)\citenamefont{Nojiri, Odintsov, and
  Faraoni}}]{Nojiri_2022}
\bibinfo{author}{\bibfnamefont{S.}~\bibnamefont{Nojiri}},
  \bibinfo{author}{\bibfnamefont{S.~D.} \bibnamefont{Odintsov}},
  \bibnamefont{and} \bibinfo{author}{\bibfnamefont{V.}~\bibnamefont{Faraoni}},
  \bibinfo{journal}{Astrophysics} \textbf{\bibinfo{volume}{65}},
  \bibinfo{pages}{534–551} (\bibinfo{year}{2022}), ISSN
  \bibinfo{issn}{1573-8191},
  \urlprefix\url{http://dx.doi.org/10.1007/s10511-023-09759-1}.

\bibitem[{\citenamefont{Manoharan et~al.}(2023)\citenamefont{Manoharan, Shaji,
  and Mathew}}]{Manoharan:2022qll}
\bibinfo{author}{\bibfnamefont{M.~T.} \bibnamefont{Manoharan}},
  \bibinfo{author}{\bibfnamefont{N.}~\bibnamefont{Shaji}}, \bibnamefont{and}
  \bibinfo{author}{\bibfnamefont{T.~K.} \bibnamefont{Mathew}},
  \bibinfo{journal}{Eur. Phys. J. C} \textbf{\bibinfo{volume}{83}},
  \bibinfo{pages}{19} (\bibinfo{year}{2023}), \eprint{2208.08736}.

\bibitem[{\citenamefont{Di~Gennaro and Ong}(2022)}]{Di_Gennaro_2022}
\bibinfo{author}{\bibfnamefont{S.}~\bibnamefont{Di~Gennaro}} \bibnamefont{and}
  \bibinfo{author}{\bibfnamefont{Y.~C.} \bibnamefont{Ong}},
  \bibinfo{journal}{Universe} \textbf{\bibinfo{volume}{8}},
  \bibinfo{pages}{541} (\bibinfo{year}{2022}), ISSN \bibinfo{issn}{2218-1997},
  \urlprefix\url{http://dx.doi.org/10.3390/universe8100541}.

\bibitem[{\citenamefont{Bhattacharjee}(2021)}]{Bhattacharjee_2021}
\bibinfo{author}{\bibfnamefont{S.}~\bibnamefont{Bhattacharjee}},
  \bibinfo{journal}{The European Physical Journal C}
  \textbf{\bibinfo{volume}{81}} (\bibinfo{year}{2021}), ISSN
  \bibinfo{issn}{1434-6052},
  \urlprefix\url{http://dx.doi.org/10.1140/epjc/s10052-021-09003-0}.

\bibitem[{\citenamefont{Moradpour et~al.}(2020)\citenamefont{Moradpour, Ziaie,
  and Kord~Zangeneh}}]{Moradpour:2020dfm}
\bibinfo{author}{\bibfnamefont{H.}~\bibnamefont{Moradpour}},
  \bibinfo{author}{\bibfnamefont{A.~H.} \bibnamefont{Ziaie}}, \bibnamefont{and}
  \bibinfo{author}{\bibfnamefont{M.}~\bibnamefont{Kord~Zangeneh}},
  \bibinfo{journal}{Eur. Phys. J. C} \textbf{\bibinfo{volume}{80}},
  \bibinfo{pages}{732} (\bibinfo{year}{2020}), \eprint{2005.06271}.

\bibitem[{\citenamefont{D'Agostino}(2019)}]{DAgostino:2019wko}
\bibinfo{author}{\bibfnamefont{R.}~\bibnamefont{D'Agostino}},
  \bibinfo{journal}{Phys. Rev. D} \textbf{\bibinfo{volume}{99}},
  \bibinfo{pages}{103524} (\bibinfo{year}{2019}), \eprint{1903.03836}.

\bibitem[{\citenamefont{Saridakis et~al.}(2018)\citenamefont{Saridakis, Bamba,
  Myrzakulov, and Anagnostopoulos}}]{Saridakis_2018}
\bibinfo{author}{\bibfnamefont{E.~N.} \bibnamefont{Saridakis}},
  \bibinfo{author}{\bibfnamefont{K.}~\bibnamefont{Bamba}},
  \bibinfo{author}{\bibfnamefont{R.}~\bibnamefont{Myrzakulov}},
  \bibnamefont{and} \bibinfo{author}{\bibfnamefont{F.~K.}
  \bibnamefont{Anagnostopoulos}}, \bibinfo{journal}{Journal of Cosmology and
  Astroparticle Physics} \textbf{\bibinfo{volume}{2018}},
  \bibinfo{pages}{012–012} (\bibinfo{year}{2018}), ISSN
  \bibinfo{issn}{1475-7516},
  \urlprefix\url{http://dx.doi.org/10.1088/1475-7516/2018/12/012}.

\bibitem[{\citenamefont{Odintsov et~al.}(2023)\citenamefont{Odintsov,
  D’Onofrio, and Paul}}]{Odintsov_2023}
\bibinfo{author}{\bibfnamefont{S.~D.} \bibnamefont{Odintsov}},
  \bibinfo{author}{\bibfnamefont{S.}~\bibnamefont{D’Onofrio}},
  \bibnamefont{and} \bibinfo{author}{\bibfnamefont{T.}~\bibnamefont{Paul}},
  \bibinfo{journal}{Physics of the Dark Universe}
  \textbf{\bibinfo{volume}{42}}, \bibinfo{pages}{101277}
  (\bibinfo{year}{2023}), ISSN \bibinfo{issn}{2212-6864},
  \urlprefix\url{http://dx.doi.org/10.1016/j.dark.2023.101277}.

\bibitem[{\citenamefont{Odintsov et~al.}(2024)\citenamefont{Odintsov,
  D'Onofrio, and Paul}}]{Odintsov:2023rqf}
\bibinfo{author}{\bibfnamefont{S.~D.} \bibnamefont{Odintsov}},
  \bibinfo{author}{\bibfnamefont{S.}~\bibnamefont{D'Onofrio}},
  \bibnamefont{and} \bibinfo{author}{\bibfnamefont{T.}~\bibnamefont{Paul}},
  \bibinfo{journal}{Universe} \textbf{\bibinfo{volume}{10}}, \bibinfo{pages}{4}
  (\bibinfo{year}{2024}), \eprint{2312.13587}.

\bibitem[{\citenamefont{Lambiase et~al.}(2023)\citenamefont{Lambiase, Luciano,
  and Sheykhi}}]{Lambiase:2023ryq}
\bibinfo{author}{\bibfnamefont{G.}~\bibnamefont{Lambiase}},
  \bibinfo{author}{\bibfnamefont{G.~G.} \bibnamefont{Luciano}},
  \bibnamefont{and} \bibinfo{author}{\bibfnamefont{A.}~\bibnamefont{Sheykhi}},
  \bibinfo{journal}{Eur. Phys. J. C} \textbf{\bibinfo{volume}{83}},
  \bibinfo{pages}{936} (\bibinfo{year}{2023}), \eprint{2307.04027}.

\bibitem[{\citenamefont{Teimoori et~al.}(2024)\citenamefont{Teimoori,
  Rezazadeh, and Rostami}}]{Teimoori:2023hpv}
\bibinfo{author}{\bibfnamefont{Z.}~\bibnamefont{Teimoori}},
  \bibinfo{author}{\bibfnamefont{K.}~\bibnamefont{Rezazadeh}},
  \bibnamefont{and} \bibinfo{author}{\bibfnamefont{A.}~\bibnamefont{Rostami}},
  \bibinfo{journal}{Eur. Phys. J. C} \textbf{\bibinfo{volume}{84}},
  \bibinfo{pages}{80} (\bibinfo{year}{2024}), \eprint{2307.11437}.

\bibitem[{\citenamefont{Luciano}(2023)}]{Luciano:2023roh}
\bibinfo{author}{\bibfnamefont{G.~G.} \bibnamefont{Luciano}},
  \bibinfo{journal}{Eur. Phys. J. C} \textbf{\bibinfo{volume}{83}},
  \bibinfo{pages}{329} (\bibinfo{year}{2023}), \eprint{2301.12509}.

\bibitem[{\citenamefont{Khodam-Mohammadi}(2024)}]{Khodam-Mohammadi:2024iuo}
\bibinfo{author}{\bibfnamefont{A.}~\bibnamefont{Khodam-Mohammadi}},
  \bibinfo{journal}{Mod. Phys. Lett. A} \textbf{\bibinfo{volume}{39}},
  \bibinfo{pages}{2450146} (\bibinfo{year}{2024}), \eprint{2409.16403}.

\bibitem[{\citenamefont{Akbar and Cai}(2007)}]{Akbar:2006kj}
\bibinfo{author}{\bibfnamefont{M.}~\bibnamefont{Akbar}} \bibnamefont{and}
  \bibinfo{author}{\bibfnamefont{R.-G.} \bibnamefont{Cai}},
  \bibinfo{journal}{Phys. Rev. D} \textbf{\bibinfo{volume}{75}},
  \bibinfo{pages}{084003} (\bibinfo{year}{2007}), \eprint{hep-th/0609128}.

\bibitem[{\citenamefont{Cai and Cao}(2007)}]{Cai:2006rs}
\bibinfo{author}{\bibfnamefont{R.-G.} \bibnamefont{Cai}} \bibnamefont{and}
  \bibinfo{author}{\bibfnamefont{L.-M.} \bibnamefont{Cao}},
  \bibinfo{journal}{Phys. Rev. D} \textbf{\bibinfo{volume}{75}},
  \bibinfo{pages}{064008} (\bibinfo{year}{2007}), \eprint{gr-qc/0611071}.

\bibitem[{\citenamefont{Sheykhi
  et~al.}(2007{\natexlab{a}})\citenamefont{Sheykhi, Wang, and
  Cai}}]{Sheykhi:2007zp}
\bibinfo{author}{\bibfnamefont{A.}~\bibnamefont{Sheykhi}},
  \bibinfo{author}{\bibfnamefont{B.}~\bibnamefont{Wang}}, \bibnamefont{and}
  \bibinfo{author}{\bibfnamefont{R.-G.} \bibnamefont{Cai}},
  \bibinfo{journal}{Nucl. Phys. B} \textbf{\bibinfo{volume}{779}},
  \bibinfo{pages}{1} (\bibinfo{year}{2007}{\natexlab{a}}),
  \eprint{hep-th/0701198}.

\bibitem[{\citenamefont{Sheykhi
  et~al.}(2007{\natexlab{b}})\citenamefont{Sheykhi, Wang, and
  Cai}}]{Sheykhi:2007gi}
\bibinfo{author}{\bibfnamefont{A.}~\bibnamefont{Sheykhi}},
  \bibinfo{author}{\bibfnamefont{B.}~\bibnamefont{Wang}}, \bibnamefont{and}
  \bibinfo{author}{\bibfnamefont{R.-G.} \bibnamefont{Cai}},
  \bibinfo{journal}{Phys. Rev. D} \textbf{\bibinfo{volume}{76}},
  \bibinfo{pages}{023515} (\bibinfo{year}{2007}{\natexlab{b}}),
  \eprint{hep-th/0701261}.

\bibitem[{\citenamefont{Jamil et~al.}(2010{\natexlab{a}})\citenamefont{Jamil,
  Saridakis, and Setare}}]{Jamil:2009eb}
\bibinfo{author}{\bibfnamefont{M.}~\bibnamefont{Jamil}},
  \bibinfo{author}{\bibfnamefont{E.~N.} \bibnamefont{Saridakis}},
  \bibnamefont{and} \bibinfo{author}{\bibfnamefont{M.~R.}
  \bibnamefont{Setare}}, \bibinfo{journal}{Phys. Rev. D}
  \textbf{\bibinfo{volume}{81}}, \bibinfo{pages}{023007}
  (\bibinfo{year}{2010}{\natexlab{a}}), \eprint{0910.0822}.

\bibitem[{\citenamefont{Cai and Ohta}(2010)}]{Cai:2009ph}
\bibinfo{author}{\bibfnamefont{R.-G.} \bibnamefont{Cai}} \bibnamefont{and}
  \bibinfo{author}{\bibfnamefont{N.}~\bibnamefont{Ohta}},
  \bibinfo{journal}{Phys. Rev. D} \textbf{\bibinfo{volume}{81}},
  \bibinfo{pages}{084061} (\bibinfo{year}{2010}), \eprint{0910.2307}.

\bibitem[{\citenamefont{Wang et~al.}(2010)\citenamefont{Wang, Jing, Ding, and
  Chen}}]{Wang:2009zv}
\bibinfo{author}{\bibfnamefont{M.}~\bibnamefont{Wang}},
  \bibinfo{author}{\bibfnamefont{J.}~\bibnamefont{Jing}},
  \bibinfo{author}{\bibfnamefont{C.}~\bibnamefont{Ding}}, \bibnamefont{and}
  \bibinfo{author}{\bibfnamefont{S.}~\bibnamefont{Chen}},
  \bibinfo{journal}{Phys. Rev. D} \textbf{\bibinfo{volume}{81}},
  \bibinfo{pages}{083006} (\bibinfo{year}{2010}), \eprint{0912.4832}.

\bibitem[{\citenamefont{Jamil et~al.}(2010{\natexlab{b}})\citenamefont{Jamil,
  Saridakis, and Setare}}]{Jamil:2010di}
\bibinfo{author}{\bibfnamefont{M.}~\bibnamefont{Jamil}},
  \bibinfo{author}{\bibfnamefont{E.~N.} \bibnamefont{Saridakis}},
  \bibnamefont{and} \bibinfo{author}{\bibfnamefont{M.~R.}
  \bibnamefont{Setare}}, \bibinfo{journal}{JCAP} \textbf{\bibinfo{volume}{11}},
  \bibinfo{pages}{032} (\bibinfo{year}{2010}{\natexlab{b}}),
  \eprint{1003.0876}.

\bibitem[{\citenamefont{Gim et~al.}(2014)\citenamefont{Gim, Kim, and
  Yi}}]{Gim:2014nba}
\bibinfo{author}{\bibfnamefont{Y.}~\bibnamefont{Gim}},
  \bibinfo{author}{\bibfnamefont{W.}~\bibnamefont{Kim}}, \bibnamefont{and}
  \bibinfo{author}{\bibfnamefont{S.-H.} \bibnamefont{Yi}},
  \bibinfo{journal}{JHEP} \textbf{\bibinfo{volume}{07}}, \bibinfo{pages}{002}
  (\bibinfo{year}{2014}), \eprint{1403.4704}.

\bibitem[{\citenamefont{Fan and Lu}(2015)}]{Fan:2014ala}
\bibinfo{author}{\bibfnamefont{Z.-Y.} \bibnamefont{Fan}} \bibnamefont{and}
  \bibinfo{author}{\bibfnamefont{H.}~\bibnamefont{Lu}}, \bibinfo{journal}{Phys.
  Rev. D} \textbf{\bibinfo{volume}{91}}, \bibinfo{pages}{064009}
  (\bibinfo{year}{2015}), \eprint{1501.00006}.

\bibitem[{\citenamefont{Sanchez and Quevedo}(2023)}]{Sanchez:2022xfh}
\bibinfo{author}{\bibfnamefont{L.~M.} \bibnamefont{Sanchez}} \bibnamefont{and}
  \bibinfo{author}{\bibfnamefont{H.}~\bibnamefont{Quevedo}},
  \bibinfo{journal}{Phys. Lett. B} \textbf{\bibinfo{volume}{839}},
  \bibinfo{pages}{137778} (\bibinfo{year}{2023}), \eprint{2208.05729}.

\bibitem[{\citenamefont{Sheykhi}(2018{\natexlab{a}})}]{Sheykhi:2018dpn}
\bibinfo{author}{\bibfnamefont{A.}~\bibnamefont{Sheykhi}},
  \bibinfo{journal}{Phys. Lett. B} \textbf{\bibinfo{volume}{785}},
  \bibinfo{pages}{118} (\bibinfo{year}{2018}{\natexlab{a}}).

\bibitem[{\citenamefont{Saridakis et~al.}(2020)\citenamefont{Saridakis,
  Basilakos, Leon, and Pavlović}}]{Saridakis:2020cqq}
\bibinfo{author}{\bibfnamefont{E.~N.} \bibnamefont{Saridakis}},
  \bibinfo{author}{\bibfnamefont{S.}~\bibnamefont{Basilakos}},
  \bibinfo{author}{\bibfnamefont{G.}~\bibnamefont{Leon}}, \bibnamefont{and}
  \bibinfo{author}{\bibfnamefont{P.}~\bibnamefont{Pavlović}},
  \bibinfo{journal}{JCAP} \textbf{\bibinfo{volume}{10}}, \bibinfo{pages}{014}
  (\bibinfo{year}{2020}).

\bibitem[{\citenamefont{Nojiri et~al.}(2017)\citenamefont{Nojiri, Odintsov, and
  Oikonomou}}]{Nojiri:2019skr}
\bibinfo{author}{\bibfnamefont{S.}~\bibnamefont{Nojiri}},
  \bibinfo{author}{\bibfnamefont{S.~D.} \bibnamefont{Odintsov}},
  \bibnamefont{and}
  \bibinfo{author}{\bibfnamefont{V.}~\bibnamefont{Oikonomou}},
  \bibinfo{journal}{Phys. Rept.} \textbf{\bibinfo{volume}{692}},
  \bibinfo{pages}{1} (\bibinfo{year}{2017}).

\bibitem[{\citenamefont{Tsallis}(1988)}]{Tsallis:1987eu}
\bibinfo{author}{\bibfnamefont{C.}~\bibnamefont{Tsallis}}, \bibinfo{journal}{J.
  Statist. Phys.} \textbf{\bibinfo{volume}{52}}, \bibinfo{pages}{479}
  (\bibinfo{year}{1988}).

\bibitem[{\citenamefont{Lyra and Tsallis}(1998)}]{Lyra:1997ggy}
\bibinfo{author}{\bibfnamefont{M.~L.} \bibnamefont{Lyra}} \bibnamefont{and}
  \bibinfo{author}{\bibfnamefont{C.}~\bibnamefont{Tsallis}},
  \bibinfo{journal}{Phys. Rev. Lett.} \textbf{\bibinfo{volume}{80}},
  \bibinfo{pages}{53} (\bibinfo{year}{1998}).

\bibitem[{\citenamefont{R\'{e}nyi}(1960)}]{Renyi:1960}
\bibinfo{author}{\bibfnamefont{A.}~\bibnamefont{R\'{e}nyi}},
  \bibinfo{journal}{Proceedings of the 4th Berkeley Symposium on Mathematics,
  Statistics and Probability, University of California Press, Berkeley and Los
  Angeles} \textbf{\bibinfo{volume}{I}}, \bibinfo{pages}{547}
  (\bibinfo{year}{1960}).

\bibitem[{\citenamefont{Kaniadakis}(2005)}]{Kaniadakis:2005zk}
\bibinfo{author}{\bibfnamefont{G.}~\bibnamefont{Kaniadakis}},
  \bibinfo{journal}{Phys. Rev. E} \textbf{\bibinfo{volume}{72}},
  \bibinfo{pages}{036108} (\bibinfo{year}{2005}), \eprint{cond-mat/0507311}.

\bibitem[{\citenamefont{Drepanou et~al.}(2022)\citenamefont{Drepanou, Lymperis,
  Saridakis, and Yesmakhanova}}]{Drepanou:2021jiv}
\bibinfo{author}{\bibfnamefont{N.}~\bibnamefont{Drepanou}},
  \bibinfo{author}{\bibfnamefont{A.}~\bibnamefont{Lymperis}},
  \bibinfo{author}{\bibfnamefont{E.~N.} \bibnamefont{Saridakis}},
  \bibnamefont{and}
  \bibinfo{author}{\bibfnamefont{K.}~\bibnamefont{Yesmakhanova}},
  \bibinfo{journal}{Eur. Phys. J. C} \textbf{\bibinfo{volume}{82}},
  \bibinfo{pages}{449} (\bibinfo{year}{2022}), \eprint{2109.09181}.

\bibitem[{\citenamefont{Jawad and Mohsaneen}(2018)}]{Jawad:2018frx}
\bibinfo{author}{\bibfnamefont{A.}~\bibnamefont{Jawad}} \bibnamefont{and}
  \bibinfo{author}{\bibfnamefont{S.}~\bibnamefont{Mohsaneen}},
  \bibinfo{journal}{Eur. Phys. J. Plus} \textbf{\bibinfo{volume}{133}},
  \bibinfo{pages}{511} (\bibinfo{year}{2018}).

\bibitem[{\citenamefont{Tavayef et~al.}(2018)\citenamefont{Tavayef, Sheykhi,
  Bamba, and Moradpour}}]{Tavayef:2018xwx}
\bibinfo{author}{\bibfnamefont{M.}~\bibnamefont{Tavayef}},
  \bibinfo{author}{\bibfnamefont{A.}~\bibnamefont{Sheykhi}},
  \bibinfo{author}{\bibfnamefont{K.}~\bibnamefont{Bamba}}, \bibnamefont{and}
  \bibinfo{author}{\bibfnamefont{H.}~\bibnamefont{Moradpour}},
  \bibinfo{journal}{Phys. Lett. B} \textbf{\bibinfo{volume}{781}},
  \bibinfo{pages}{195} (\bibinfo{year}{2018}).

\bibitem[{\citenamefont{Moradpour et~al.}(2017)\citenamefont{Moradpour,
  Sadeghnezhad, and Hendi}}]{Moradpour:2016ubd}
\bibinfo{author}{\bibfnamefont{H.}~\bibnamefont{Moradpour}},
  \bibinfo{author}{\bibfnamefont{N.}~\bibnamefont{Sadeghnezhad}},
  \bibnamefont{and} \bibinfo{author}{\bibfnamefont{S.~H.} \bibnamefont{Hendi}},
  \bibinfo{journal}{Can. J. Phys.} \textbf{\bibinfo{volume}{95}},
  \bibinfo{pages}{1257} (\bibinfo{year}{2017}).

\bibitem[{\citenamefont{Luciano and Petruzziello}(2019)}]{Luciano:2019cna}
\bibinfo{author}{\bibfnamefont{G.~G.} \bibnamefont{Luciano}} \bibnamefont{and}
  \bibinfo{author}{\bibfnamefont{L.}~\bibnamefont{Petruzziello}},
  \bibinfo{journal}{Eur. Phys. J. C} \textbf{\bibinfo{volume}{79}},
  \bibinfo{pages}{283} (\bibinfo{year}{2019}).

\bibitem[{\citenamefont{Ghaffari and Nozari}(2022)}]{Ghaffari:2022tiq}
\bibinfo{author}{\bibfnamefont{S.}~\bibnamefont{Ghaffari}} \bibnamefont{and}
  \bibinfo{author}{\bibfnamefont{K.}~\bibnamefont{Nozari}},
  \bibinfo{journal}{Eur. Phys. J. C} \textbf{\bibinfo{volume}{82}},
  \bibinfo{pages}{1056} (\bibinfo{year}{2022}).

\bibitem[{\citenamefont{Cai and Kim}(2005)}]{Cai:2005ra}
\bibinfo{author}{\bibfnamefont{R.-G.} \bibnamefont{Cai}} \bibnamefont{and}
  \bibinfo{author}{\bibfnamefont{S.~P.} \bibnamefont{Kim}},
  \bibinfo{journal}{JHEP} \textbf{\bibinfo{volume}{0502}}, \bibinfo{pages}{050}
  (\bibinfo{year}{2005}).

\bibitem[{\citenamefont{Sheykhi}(2018{\natexlab{b}})}]{Sheykhi_2018}
\bibinfo{author}{\bibfnamefont{A.}~\bibnamefont{Sheykhi}},
  \bibinfo{journal}{Physics Letters B} \textbf{\bibinfo{volume}{785}},
  \bibinfo{pages}{118–126} (\bibinfo{year}{2018}{\natexlab{b}}), ISSN
  \bibinfo{issn}{0370-2693},
  \urlprefix\url{http://dx.doi.org/10.1016/j.physletb.2018.08.036}.

\bibitem[{\citenamefont{Odintsov and Paul}(2023)}]{Odintsov:2022qnn}
\bibinfo{author}{\bibfnamefont{S.~D.} \bibnamefont{Odintsov}} \bibnamefont{and}
  \bibinfo{author}{\bibfnamefont{T.}~\bibnamefont{Paul}},
  \bibinfo{journal}{Phys. Dark Univ.} \textbf{\bibinfo{volume}{39}},
  \bibinfo{pages}{101159} (\bibinfo{year}{2023}), \eprint{2212.05531}.

\bibitem[{\citenamefont{Odintsov and Oikonomou}(2020)}]{Odintsov:2020zct}
\bibinfo{author}{\bibfnamefont{S.~D.} \bibnamefont{Odintsov}} \bibnamefont{and}
  \bibinfo{author}{\bibfnamefont{V.~K.} \bibnamefont{Oikonomou}},
  \bibinfo{journal}{Phys. Rev. D} \textbf{\bibinfo{volume}{101}},
  \bibinfo{pages}{044009} (\bibinfo{year}{2020}).

\bibitem[{\citenamefont{Khodam-Mohammadi and
  Monshizadeh}(2023)}]{Khodam_Mohammadi_2023}
\bibinfo{author}{\bibfnamefont{A.}~\bibnamefont{Khodam-Mohammadi}}
  \bibnamefont{and}
  \bibinfo{author}{\bibfnamefont{M.}~\bibnamefont{Monshizadeh}},
  \bibinfo{journal}{Physics Letters B} \textbf{\bibinfo{volume}{843}},
  \bibinfo{pages}{138066} (\bibinfo{year}{2023}), ISSN
  \bibinfo{issn}{0370-2693},
  \urlprefix\url{http://dx.doi.org/10.1016/j.physletb.2023.138066}.

\bibitem[{\citenamefont{Scolnic et~al.}(2025)\citenamefont{Scolnic, Riess,
  Murakami, Peterson, Brout, Acevedo, Carreres, Jones, Said, Howlett
  et~al.}}]{scolnic2025hubble}
\bibinfo{author}{\bibfnamefont{D.}~\bibnamefont{Scolnic}},
  \bibinfo{author}{\bibfnamefont{A.~G.} \bibnamefont{Riess}},
  \bibinfo{author}{\bibfnamefont{Y.~S.} \bibnamefont{Murakami}},
  \bibinfo{author}{\bibfnamefont{E.~R.} \bibnamefont{Peterson}},
  \bibinfo{author}{\bibfnamefont{D.}~\bibnamefont{Brout}},
  \bibinfo{author}{\bibfnamefont{M.}~\bibnamefont{Acevedo}},
  \bibinfo{author}{\bibfnamefont{B.}~\bibnamefont{Carreres}},
  \bibinfo{author}{\bibfnamefont{D.~O.} \bibnamefont{Jones}},
  \bibinfo{author}{\bibfnamefont{K.}~\bibnamefont{Said}},
  \bibinfo{author}{\bibfnamefont{C.}~\bibnamefont{Howlett}},
  \bibnamefont{et~al.}, \bibinfo{journal}{The Astrophysical Journal Letters}
  \textbf{\bibinfo{volume}{979}}, \bibinfo{pages}{L9} (\bibinfo{year}{2025}).

\bibitem[{\citenamefont{Riess et~al.}(2024)\citenamefont{Riess, Anand, Yuan,
  Casertano, Dolphin, Macri, Breuval, Scolnic, Perrin, and
  Anderson}}]{riess2024jwst}
\bibinfo{author}{\bibfnamefont{A.~G.} \bibnamefont{Riess}},
  \bibinfo{author}{\bibfnamefont{G.~S.} \bibnamefont{Anand}},
  \bibinfo{author}{\bibfnamefont{W.}~\bibnamefont{Yuan}},
  \bibinfo{author}{\bibfnamefont{S.}~\bibnamefont{Casertano}},
  \bibinfo{author}{\bibfnamefont{A.}~\bibnamefont{Dolphin}},
  \bibinfo{author}{\bibfnamefont{L.~M.} \bibnamefont{Macri}},
  \bibinfo{author}{\bibfnamefont{L.}~\bibnamefont{Breuval}},
  \bibinfo{author}{\bibfnamefont{D.}~\bibnamefont{Scolnic}},
  \bibinfo{author}{\bibfnamefont{M.}~\bibnamefont{Perrin}}, \bibnamefont{and}
  \bibinfo{author}{\bibfnamefont{R.~I.} \bibnamefont{Anderson}},
  \bibinfo{journal}{The Astrophysical Journal Letters}
  \textbf{\bibinfo{volume}{962}}, \bibinfo{pages}{L17} (\bibinfo{year}{2024}).

\bibitem[{\citenamefont{Hart and Chluba}(2020)}]{hart2020updated}
\bibinfo{author}{\bibfnamefont{L.}~\bibnamefont{Hart}} \bibnamefont{and}
  \bibinfo{author}{\bibfnamefont{J.}~\bibnamefont{Chluba}},
  \bibinfo{journal}{Monthly Notices of the Royal Astronomical Society}
  \textbf{\bibinfo{volume}{493}}, \bibinfo{pages}{3255} (\bibinfo{year}{2020}).

\bibitem[{\citenamefont{Vacher et~al.}(2022)\citenamefont{Vacher, Dias,
  Sch{\"o}neberg, Martins, Vinzl, Nesseris, Canas-Herrera, and
  Martinelli}}]{vacher2022constraints}
\bibinfo{author}{\bibfnamefont{L.}~\bibnamefont{Vacher}},
  \bibinfo{author}{\bibfnamefont{J.}~\bibnamefont{Dias}},
  \bibinfo{author}{\bibfnamefont{N.}~\bibnamefont{Sch{\"o}neberg}},
  \bibinfo{author}{\bibfnamefont{C.}~\bibnamefont{Martins}},
  \bibinfo{author}{\bibfnamefont{S.}~\bibnamefont{Vinzl}},
  \bibinfo{author}{\bibfnamefont{S.}~\bibnamefont{Nesseris}},
  \bibinfo{author}{\bibfnamefont{G.}~\bibnamefont{Canas-Herrera}},
  \bibnamefont{and}
  \bibinfo{author}{\bibfnamefont{M.}~\bibnamefont{Martinelli}},
  \bibinfo{journal}{Physical Review D} \textbf{\bibinfo{volume}{106}},
  \bibinfo{pages}{083522} (\bibinfo{year}{2022}).

\bibitem[{\citenamefont{Tyagi et~al.}(2025)\citenamefont{Tyagi, Haridasu, and
  Basak}}]{tyagi2025constraints}
\bibinfo{author}{\bibfnamefont{U.~K.} \bibnamefont{Tyagi}},
  \bibinfo{author}{\bibfnamefont{S.}~\bibnamefont{Haridasu}}, \bibnamefont{and}
  \bibinfo{author}{\bibfnamefont{S.}~\bibnamefont{Basak}},
  \bibinfo{journal}{arXiv preprint arXiv:2504.11308}  (\bibinfo{year}{2025}).

\bibitem[{\citenamefont{Cattoen and Visser}(2007)}]{Cattoen:2007sk}
\bibinfo{author}{\bibfnamefont{C.}~\bibnamefont{Cattoen}} \bibnamefont{and}
  \bibinfo{author}{\bibfnamefont{M.}~\bibnamefont{Visser}},
  \bibinfo{journal}{Class. Quant. Grav.} \textbf{\bibinfo{volume}{24}},
  \bibinfo{pages}{5985} (\bibinfo{year}{2007}), \eprint{0710.1887}.

\bibitem[{\citenamefont{Capozziello et~al.}(2011)\citenamefont{Capozziello,
  Lazkoz, and Salzano}}]{Capozziello:2011tj}
\bibinfo{author}{\bibfnamefont{S.}~\bibnamefont{Capozziello}},
  \bibinfo{author}{\bibfnamefont{R.}~\bibnamefont{Lazkoz}}, \bibnamefont{and}
  \bibinfo{author}{\bibfnamefont{V.}~\bibnamefont{Salzano}},
  \bibinfo{journal}{Phys. Rev. D} \textbf{\bibinfo{volume}{84}},
  \bibinfo{pages}{124061} (\bibinfo{year}{2011}), \eprint{1104.3096}.

\bibitem[{\citenamefont{Rezaei et~al.}(2020)\citenamefont{Rezaei, Pour-Ojaghi,
  and Malekjani}}]{Rezaei:2020lfy}
\bibinfo{author}{\bibfnamefont{M.}~\bibnamefont{Rezaei}},
  \bibinfo{author}{\bibfnamefont{S.}~\bibnamefont{Pour-Ojaghi}},
  \bibnamefont{and}
  \bibinfo{author}{\bibfnamefont{M.}~\bibnamefont{Malekjani}},
  \bibinfo{journal}{Astrophys. J.} \textbf{\bibinfo{volume}{900}},
  \bibinfo{pages}{70} (\bibinfo{year}{2020}), \eprint{2008.03092}.

\bibitem[{\citenamefont{Li et~al.}(2020)\citenamefont{Li, Du, and
  Xu}}]{Li:2019qic}
\bibinfo{author}{\bibfnamefont{E.-K.} \bibnamefont{Li}},
  \bibinfo{author}{\bibfnamefont{M.}~\bibnamefont{Du}}, \bibnamefont{and}
  \bibinfo{author}{\bibfnamefont{L.}~\bibnamefont{Xu}}, \bibinfo{journal}{Mon.
  Not. Roy. Astron. Soc.} \textbf{\bibinfo{volume}{491}}, \bibinfo{pages}{4960}
  (\bibinfo{year}{2020}), \eprint{1903.11433}.

\bibitem[{\citenamefont{Capozziello et~al.}(2020)\citenamefont{Capozziello,
  D'Agostino, and Luongo}}]{Capozziello:2020ctn}
\bibinfo{author}{\bibfnamefont{S.}~\bibnamefont{Capozziello}},
  \bibinfo{author}{\bibfnamefont{R.}~\bibnamefont{D'Agostino}},
  \bibnamefont{and} \bibinfo{author}{\bibfnamefont{O.}~\bibnamefont{Luongo}},
  \bibinfo{journal}{Mon. Not. Roy. Astron. Soc.}
  \textbf{\bibinfo{volume}{494}}, \bibinfo{pages}{2576} (\bibinfo{year}{2020}),
  \eprint{2003.09341}.

\bibitem[{\citenamefont{Vitagliano et~al.}(2010)\citenamefont{Vitagliano, Xia,
  Liberati, and Viel}}]{Vitagliano:2009et}
\bibinfo{author}{\bibfnamefont{V.}~\bibnamefont{Vitagliano}},
  \bibinfo{author}{\bibfnamefont{J.-Q.} \bibnamefont{Xia}},
  \bibinfo{author}{\bibfnamefont{S.}~\bibnamefont{Liberati}}, \bibnamefont{and}
  \bibinfo{author}{\bibfnamefont{M.}~\bibnamefont{Viel}},
  \bibinfo{journal}{JCAP} \textbf{\bibinfo{volume}{03}}, \bibinfo{pages}{005}
  (\bibinfo{year}{2010}), \eprint{0911.1249}.

\bibitem[{\citenamefont{Shahhoseini
  et~al.}(2025{\natexlab{b}})\citenamefont{Shahhoseini, Malekjani, and
  Khodam-Mohammadi}}]{Shahhoseini:2025pyr}
\bibinfo{author}{\bibfnamefont{N.}~\bibnamefont{Shahhoseini}},
  \bibinfo{author}{\bibfnamefont{M.}~\bibnamefont{Malekjani}},
  \bibnamefont{and}
  \bibinfo{author}{\bibfnamefont{A.}~\bibnamefont{Khodam-Mohammadi}},
  \bibinfo{journal}{Eur. Phys. J. C} \textbf{\bibinfo{volume}{85}},
  \bibinfo{pages}{699} (\bibinfo{year}{2025}{\natexlab{b}}).

\bibitem[{\citenamefont{Rezaei et~al.}(2017)\citenamefont{Rezaei, Malekjani,
  Basilakos, Mehrabi, and Mota}}]{rezaei2017constraints}
\bibinfo{author}{\bibfnamefont{M.}~\bibnamefont{Rezaei}},
  \bibinfo{author}{\bibfnamefont{M.}~\bibnamefont{Malekjani}},
  \bibinfo{author}{\bibfnamefont{S.}~\bibnamefont{Basilakos}},
  \bibinfo{author}{\bibfnamefont{A.}~\bibnamefont{Mehrabi}}, \bibnamefont{and}
  \bibinfo{author}{\bibfnamefont{D.~F.} \bibnamefont{Mota}},
  \bibinfo{journal}{The Astrophysical Journal} \textbf{\bibinfo{volume}{843}},
  \bibinfo{pages}{65} (\bibinfo{year}{2017}).

\bibitem[{\citenamefont{Pourojaghi and Malekjani}(2021)}]{Pourojaghi:2021den}
\bibinfo{author}{\bibfnamefont{S.}~\bibnamefont{Pourojaghi}} \bibnamefont{and}
  \bibinfo{author}{\bibfnamefont{M.}~\bibnamefont{Malekjani}},
  \bibinfo{journal}{Eur. Phys. J. C} \textbf{\bibinfo{volume}{81}},
  \bibinfo{pages}{575} (\bibinfo{year}{2021}).

\bibitem[{\citenamefont{Visser}(2004)}]{Visser:2003vq}
\bibinfo{author}{\bibfnamefont{M.}~\bibnamefont{Visser}},
  \bibinfo{journal}{Class. Quant. Grav.} \textbf{\bibinfo{volume}{21}},
  \bibinfo{pages}{2603} (\bibinfo{year}{2004}), \eprint{gr-qc/0309109}.

\bibitem[{\citenamefont{Pourojaghi et~al.}(2022)\citenamefont{Pourojaghi,
  Zabihi, and Malekjani}}]{Pourojaghi:2022zrh}
\bibinfo{author}{\bibfnamefont{S.}~\bibnamefont{Pourojaghi}},
  \bibinfo{author}{\bibfnamefont{N.~F.} \bibnamefont{Zabihi}},
  \bibnamefont{and}
  \bibinfo{author}{\bibfnamefont{M.}~\bibnamefont{Malekjani}},
  \bibinfo{journal}{Phys. Rev. D} \textbf{\bibinfo{volume}{106}},
  \bibinfo{pages}{123523} (\bibinfo{year}{2022}), \eprint{2212.04118}.

\bibitem[{\citenamefont{Khodam-Mohammadi and
  Jafari}(2023)}]{Khodam-Mohammadi:2023PLB}
\bibinfo{author}{\bibfnamefont{A.}~\bibnamefont{Khodam-Mohammadi}}
  \bibnamefont{and} \bibinfo{author}{\bibfnamefont{F.}~\bibnamefont{Jafari}},
  \bibinfo{journal}{Phys. Lett. B} \textbf{\bibinfo{volume}{839}},
  \bibinfo{pages}{137760} (\bibinfo{year}{2023}).

\bibitem[{\citenamefont{Damour and Esposito-Farese}(1996)}]{Damour:1996ke}
\bibinfo{author}{\bibfnamefont{T.}~\bibnamefont{Damour}} \bibnamefont{and}
  \bibinfo{author}{\bibfnamefont{G.}~\bibnamefont{Esposito-Farese}},
  \bibinfo{journal}{Phys. Rev. D} \textbf{\bibinfo{volume}{54}},
  \bibinfo{pages}{1474} (\bibinfo{year}{1996}), \eprint{gr-qc/9602056}.

\bibitem[{\citenamefont{Fujii and Maeda}(2007)}]{Fujii:2003pa}
\bibinfo{author}{\bibfnamefont{Y.}~\bibnamefont{Fujii}} \bibnamefont{and}
  \bibinfo{author}{\bibfnamefont{K.}~\bibnamefont{Maeda}},
  \emph{\bibinfo{title}{{The scalar-tensor theory of gravitation}}}, Cambridge
  Monographs on Mathematical Physics (\bibinfo{publisher}{Cambridge University
  Press}, \bibinfo{year}{2007}), ISBN \bibinfo{isbn}{978-0-521-03752-5,
  978-0-521-81159-0, 978-0-511-02988-2}.

\bibitem[{\citenamefont{Faraoni}(2004)}]{Faraoni:2004pi}
\bibinfo{author}{\bibfnamefont{V.}~\bibnamefont{Faraoni}},
  \emph{\bibinfo{title}{{Cosmology in scalar tensor gravity}}}
  (\bibinfo{year}{2004}), ISBN \bibinfo{isbn}{978-1-4020-1988-3}.

\bibitem[{\citenamefont{Prigogine et~al.}(1989)\citenamefont{Prigogine,
  Geheniau, Gunzig, and Nardone}}]{Prigogine:1989zz}
\bibinfo{author}{\bibfnamefont{I.}~\bibnamefont{Prigogine}},
  \bibinfo{author}{\bibfnamefont{J.}~\bibnamefont{Geheniau}},
  \bibinfo{author}{\bibfnamefont{E.}~\bibnamefont{Gunzig}}, \bibnamefont{and}
  \bibinfo{author}{\bibfnamefont{P.}~\bibnamefont{Nardone}},
  \bibinfo{journal}{Gen. Rel. Grav.} \textbf{\bibinfo{volume}{21}},
  \bibinfo{pages}{767} (\bibinfo{year}{1989}).

\bibitem[{\citenamefont{Lima et~al.}(1996)\citenamefont{Lima, Germano, and
  Abramo}}]{Lima:1995xz}
\bibinfo{author}{\bibfnamefont{J.~A.~S.} \bibnamefont{Lima}},
  \bibinfo{author}{\bibfnamefont{A.~S.~M.} \bibnamefont{Germano}},
  \bibnamefont{and} \bibinfo{author}{\bibfnamefont{L.~R.~W.}
  \bibnamefont{Abramo}}, \bibinfo{journal}{Phys. Rev. D}
  \textbf{\bibinfo{volume}{53}}, \bibinfo{pages}{4287} (\bibinfo{year}{1996}),
  \eprint{gr-qc/9511006}.

\bibitem[{\citenamefont{Pan et~al.}(2018)\citenamefont{Pan, Saridakis, and
  Yang}}]{Pan:2017zoh}
\bibinfo{author}{\bibfnamefont{S.}~\bibnamefont{Pan}},
  \bibinfo{author}{\bibfnamefont{E.~N.} \bibnamefont{Saridakis}},
  \bibnamefont{and} \bibinfo{author}{\bibfnamefont{W.}~\bibnamefont{Yang}},
  \bibinfo{journal}{Phys. Rev. D} \textbf{\bibinfo{volume}{98}},
  \bibinfo{pages}{063510} (\bibinfo{year}{2018}), \eprint{1712.05746}.

\bibitem[{\citenamefont{Sola}(2013)}]{Sola:2013gha}
\bibinfo{author}{\bibfnamefont{J.}~\bibnamefont{Sola}}, \bibinfo{journal}{J.
  Phys. Conf. Ser.} \textbf{\bibinfo{volume}{453}}, \bibinfo{pages}{012015}
  (\bibinfo{year}{2013}), \eprint{1306.1527}.

\bibitem[{\citenamefont{Wetterich}(2014)}]{Wetterich:2013jsa}
\bibinfo{author}{\bibfnamefont{C.}~\bibnamefont{Wetterich}},
  \bibinfo{journal}{Phys. Rev. D} \textbf{\bibinfo{volume}{89}},
  \bibinfo{pages}{024005} (\bibinfo{year}{2014}), \eprint{1308.1019}.

\bibitem[{\citenamefont{Moradpour et~al.}(2018)\citenamefont{Moradpour,
  Moosavi, Lobo, Morais~Gra{\c{c}}a, Jawad, and Salako}}]{Moradpour:2018ivi}
\bibinfo{author}{\bibfnamefont{H.}~\bibnamefont{Moradpour}},
  \bibinfo{author}{\bibfnamefont{S.~A.} \bibnamefont{Moosavi}},
  \bibinfo{author}{\bibfnamefont{I.~P.} \bibnamefont{Lobo}},
  \bibinfo{author}{\bibfnamefont{J.~P.} \bibnamefont{Morais~Gra{\c{c}}a}},
  \bibinfo{author}{\bibfnamefont{A.}~\bibnamefont{Jawad}}, \bibnamefont{and}
  \bibinfo{author}{\bibfnamefont{I.~G.} \bibnamefont{Salako}},
  \bibinfo{journal}{Eur. Phys. J. C} \textbf{\bibinfo{volume}{78}},
  \bibinfo{pages}{829} (\bibinfo{year}{2018}), \eprint{1803.02195}.

\end{thebibliography}

\end{document}